\documentclass[12pt]{article}

\usepackage{scicite}

\usepackage{times}
\usepackage{color}
\usepackage{graphicx} 

\usepackage{amsmath}

\newenvironment{sciabstract}{%
\begin{quote} \bf}
{\end{quote}}

\newcommand{\an}[1]{\left\langle{#1}\right\rangle}
\newcounter{lastnote}

\title{Quantum-enhanced optical phase tracking}

\author{
Hidehiro Yonezawa,$^{1}$ 
Daisuke Nakane,$^{1}$ 
Trevor A. Wheatley,$^{1,2,3}$ \\
Kohjiro Iwasawa,$^1$ 
Shuntaro Takeda,$^{1}$ 
Hajime Arao,$^{1}$ 
Kentaro Ohki,$^4$ \\
Koji Tsumura,$^5 $ 
Dominic W. Berry,$^{6,7}$ 
Timothy C. Ralph,$^{2,8}$ \\ 
Howard M. Wiseman,$^{2,9\ast}$ 
Elanor H. Huntington,$^{2,3}$ 
Akira Furusawa$^{1\ast}$ \\
\\
\normalsize{$^1$Department of Applied Physics, School of Engineering,} \\
\normalsize{The University of Tokyo, 7-3-1 Hongo, Bunkyo-ku, Tokyo 113-8656, Japan} \\ 
\normalsize{$^2$Centre for Quantum Computation and Communication Technology, Australian Research Council} \\
\normalsize{$^3$School of Engineering and Information Technology, University College,} \\ 
\normalsize{The University of New South Wales, Canberra ACT 2600, Australia} \\
\normalsize{$^4$Department of Applied Mathematics and Physics, Graduate School of Informatics,} \\
\normalsize{Kyoto University, Yoshida-Honmachi, Sakyo-ku, Kyoto 606-8501, Japan} \\ 
\normalsize{$^5$Department of Information Physics and Computing,} \\
\normalsize{The University of Tokyo, 7-3-1 Hongo, Bunkyo-ku, Tokyo 113-0033, Japan} \\ 
\normalsize{$^6$Institute for Quantum Computing, University of Waterloo, Ontario N2L 3G1, Canada} \\
\normalsize{$^7$Department of Physics and Astronomy, Macquarie University, NSW 2109, Australia} \\
\normalsize{$^8$School of Mathematics and Physics, University of Queensland, Brisbane QLD 4072, Australia} \\
\normalsize{$^9$Centre for Quantum Dynamics, Griffith University, Brisbane QLD 4111, Australia } \\
\\
\normalsize{$^\ast$To whom correspondence should be addressed. E-mail:} \\ 
\normalsize{h.wiseman@griffith.edu.au (H.W.);  akiraf@ap.t.u-tokyo.ac.jp (A.F.)}
}

\date{}

\begin{document} 


\baselineskip24pt


\maketitle


\begin{sciabstract}
Tracking a randomly varying optical phase is a key task in metrology, with applications in optical communication. The best precision for optical phase tracking has till now been limited by the quantum vacuum fluctuations of coherent light. Here we surpass this coherent-state limit by using a continuous-wave beam in a phase-squeezed quantum state. Unlike in previous squeezing-enhanced metrology, restricted to phases with very small variation, the best tracking precision (for a fixed light intensity) is achieved for a finite degree of squeezing, due to Heisenberg's uncertainty principle. By optimizing the squeezing we track the phase with a mean square error 15$\pm$4\% below the coherent-state limit. 
\end{sciabstract}


There are many tasks where precise optical phase estimation is critical, including communication \cite{Slavik10,Chen12}, and metrology \cite{GLM11}. Quantum mechanics imposes a fundamental bound on precision \cite{Helstrom76,Giovannetti04,Wiseman10}, and this already limits gravitational wave detection \cite{Caves81,Goda08,LIGO2011} and can guarantee security in quantum cryptography \cite{DPS_QKD}. 
The quantum limits are determined by optimizing (subject to constraints) the input quantum state, the quantum measurement, and the data processing. Much effort has been devoted to approaching the fundamental quantum limits \cite{Giovannetti04,Wiseman10,GLM11}. 

Phase estimation can be divided into two kinds \cite{Xiang11}: phase sensing, where the phase is known to always lie within some small interval (e.g. \cite{Nagata07}), and general phase estimation, where it is not so constrained (e.g. \cite{Xiang11}). 
In the former case, when (as in most practical situations) the field has a large coherent amplitude, the problem can be linearized in terms of the phase rotation \cite{Caves81}, which greatly simplifies the task of optimizing the input state and the measurement. 
By contrast, in the case of unconstrained phase {estimation} the problem cannot be linearized, and
as a consequence the optimization problem is considerably harder \cite{Wiseman95,WisKil98,BerWis00,Armen02,Berry02,Berry06,Higgins07,Hentschel10,Xiang11,Wheatley10,Tsang09PRA}. 
While a quantum  enhancement of phase sensing using nonclassical states of light has recently been demonstrated \cite{Goda08,LIGO2011}, this has been done for general phase estimation only with post-selected results \cite{Xiang11}. 

We present a demonstration of unconstrained phase estimation with a quantum enhancement using nonclassical (squeezed) states. 
We use homodyne detection, with no post-selection of data, and no compensation for losses or detector inefficiency in the system. 
Moreover, the problem of a stochastically varying phase is addressed \cite{Goda08,LIGO2011,Berry02,Berry06,Wheatley10,Tsang09PRA}, as is highly relevant for physical metrology and communication, rather than a time-invariant (but initially unknown) phase \cite{Wiseman95,BerWis00,Armen02,Higgins07,Hentschel10,Xiang11,Nagata07}. 
To perform optimal estimation we have implemented optical phase tracking --- 
a phase-lock loop which strives to maintain the maximum measurement sensitivity for a widely varying phase. 
The quantum noise in the photocurrent prevents the maximum sensitivity from being perfectly maintained, 
which is why we observe an optimal degree of squeezing. 

Our experiment (Fig.~1A) uses a continuous-wave optical phase-squeezed beam. 
The phase of the beam is modulated with the signal $\varphi(t)$, the waveform to be estimated \cite{Tsang09PRA,Tsang09,Tsang11}. 
As in Refs.~\cite{Wheatley10,Tsang09PRA}, we use a stochastic waveform defined by 
       \begin{equation}
          \varphi(t)=\sqrt{\kappa} \int^t_{-\infty} e^{-\lambda(t-s)} dV(s). 
        \label{eq:OUnoiseM}
       \end{equation}
Here $dV(s)$ is a classically generated Wiener process~\cite{Gar85} (white noise), $\lambda^{-1}$ is the correlation time of $\varphi(t)$, and $\kappa$ determines the magnitude of the  phase variation, which is of order unity. 
This $\varphi(t)$ is a continuous-time random walk with a tendency to return to the mean phase of zero,  a kind of noisy relaxation process that occurs in many physical situations~\cite{Gar85}. 

The phase-modulated beam is measured by homodyne detection, using a local oscillator (LO), yielding a noisy current $I(t)$. 
The LO phase $\Phi(t)$ is feedback-controlled to be $\Phi(t) \approx\pi/2+\varphi(t)$, as this is the most sensitive operating point for sensing changes in $\varphi(t)$ (Fig.~1B). 
Because $\varphi(t)$ is unknown, the best strategy is adaptive metrology \cite{Wiseman95,BerWis00,Armen02,Berry02,Berry06,Higgins07,Hentschel10,Xiang11,Wheatley10}, in which feedback control is used to set $\Phi(t) = \varphi_f(t) + \pi/2$, where $\varphi_f(t)$  is a filtered estimate of $\varphi(t)$ --- that is, an estimate based on $I(s)$ for all $s<t$. 
 This gives a normalized homodyne output current $I(t)$ of \cite{Berry02,Sup},
   \begin{displaymath}
          I(t)dt  \simeq 2\left| \alpha \right | 
                      \left[ 
                                   \varphi(t)-\varphi_f(t) 
                      \right] dt 
                 +\sqrt{\bar{R}_{\rm  sq}}\, dW(t), 
   \end{displaymath}
   \begin{equation} 
        \bar{R}_{\rm  sq}  =   
                         \sigma_f^2 e^{2r_p} + (1-\sigma_f^2) e^{-2r_m}.
        \label{eq:Iapp and RsqSS}
   \end{equation}
Here $|\alpha|$ is the amplitude of the input phase-squeezed beam, and 
$dW(t)$ is another Wiener process \cite{Gar85}, arising from the squeezed vacuum fluctuations. The magnitude $\bar{R}_{\rm sq}$ of these quantum fluctuations is determined by the degree of squeezing ($r_m \geq 0$) and anti-squeezing ($r_p \geq r_m$), and by $\sigma_f^2=\left< \left[ \varphi(t)-\varphi_f(t) \right]^2\right>$. Note that several approximations --- 
justified in Ref.~\cite{Sup} ---   have been made to derive Eq.~\ref{eq:Iapp and RsqSS}, most notably, a second-order expansion for $I(t)dt$ in the small variable $\left[\varphi(t)-\varphi_f(t)\right]$. 

For optimal feedback control, the Kalman filter is used for $\varphi_f(t)$ \cite{Tsang09PRA}, 
which is the causal (i.e.~real time) estimator with the lowest mean square error (MSE). 
The Kalman filter is the optimal filter for estimating $\varphi(t)$ of the form of Eq.~\ref{eq:OUnoiseM} when using a coherent beam \cite{Tsang09PRA}, and the calculation generalizes to our squeezed case \cite{Sup}. 
Though the filtered estimate $\varphi_f(t)$ is a good estimate of the signal phase $\varphi(t)$, to obtain the best estimate we apply 
the acausal technique of smoothing \cite{Wheatley10,Tsang09PRA,Tsang09}. After storing data over a certain period of time, a precise estimate of $\varphi_s(t)$ is obtained at a time $t$ in the middle of that period by using observations both before and after $t$. 
The MSE of the smoothed estimate $\sigma_s^2=\left< \left[ \varphi(t)-\varphi_s(t) \right]^2\right>$ is given as \cite{Tsang09PRA,Sup},
      \begin{equation}
         \sigma_s^2 = {\kappa}/ {\left( 2\sqrt{
                             4\kappa \left| \alpha \right|^2/\bar{R}_{\rm  sq}
                               +\lambda^2 
                                             }\right)}.
       \label{eq:MSESM}
      \end{equation} 
Recall that $\bar{R}_{\rm  sq}$ (Eq.~\ref{eq:Iapp and RsqSS}) is a function of $\sigma_f^2$, so the above expression for 
$\sigma_{s}^{2}$ is still implicit. The full solutions are given in \cite{Sup}, but in the parameter regime of 
our experiment, $\sigma_{s}^{2}$ is approximately proportional to $\sqrt{\bar R_{\rm  sq}}$. That is, 
by using a nonclassical beam with effective squeezing $\bar R_{\rm  sq}<1$ we expect to be able to overcome the coherent-state limit (CSL)  by a factor of $\sqrt{\bar R_{\rm  sq}}$. 

Our experiment (see Fig.~1C) uses an 860 nm continuous-wave Titanium Sapphire laser. 
The phase-squeezed beam is added by an optical parametric oscillator (OPO). The OPO is driven below threshold by a 430 nm pump beam, generated by a second-harmonic-generation cavity. We obtain up to $-4$ dB of phase squeezing.
The signal $\varphi(t)$ is produced by a digital random signal generator
and a low-pass filter with a cutoff frequency of $\lambda / 2 \pi$. This is imposed upon the phase-squeezed beam using an electro-optic modulator. 
Homodyne detection is performed on this phase-modulated beam with an overall efficiency of $\eta=0.85$. 
The homodyne current goes to the optimal feedback filter (another low-pass filter with a cutoff frequency $\lambda / 2\pi$ \cite{Sup}). Its output, $\varphi_f(t)$, is then shifted by $\pi$/2 and applied on the phase of the LO beam with another electro-optic modulator. 

We record $\varphi(t)$, $I(t)$ and $\varphi_f(t)$ by an oscilloscope with a sampling rate of 100 MHz. 
Figure 2 shows a typical segment of the recorded signals, plus the smoothed estimate $\varphi_s(t)$. 
The parameters here are $\kappa= (1.9 \pm 0.1 )\times10^4$ rad/s, $\lambda= (5.9 \pm 0.5)\times10^4$ rad/s, $|\alpha|^2=(1.00 \pm 0.06)\times10^6$ s$^{-1}$, squeezing $-$3.1$\pm0.1$ dB ($r_m=0.36 \pm 0.01$) and anti-squeezing 5.1$\pm0.1$ dB ($r_p=0.59 \pm 0.01$), from a pump beam power of 80 mW. 
Note that $\kappa$ and $\lambda$ are fixed through this paper.
The current $I(t)$ has zero mean because the feedback loop is designed to operate the homodyne measurement at this point of maximum sensitivity (Fig.~1B). 
While the filtered estimate $\varphi_f(t)$ has a visible delay due to its causal nature, the smoothed estimate $\varphi_s(t)$ does not, and the signal phase $\varphi(t)$ is reliably tracked.

To investigate the squeezing-enhancement, we perform phase tracking with a fixed square amplitude $|\alpha|^2=(1.00 \pm 0.06)\times10^6$ s$^{-1}$ but with varying squeezing levels arising from OPO pump beam powers of 0, 30, 80 and 180 mW. 
Independently of the phase estimation, squeezing and anti-squeezing levels were measured for each pump beam power \cite{Sup}.
The red crosses in Fig.~3 show the MSEs of the smoothed estimates $\sigma_s^2$ as a function of the squeezing level. The MSE was calculated from 2 ms of data ($2\times10^5$ samples). 
Repeating this 15 times gave the average MSE and its standard deviation. 

Figure 3 shows three key results. 
First, the squeezing-enhancement is verified: the MSEs are reduced below the CSL (i) by using phase-squeezed beams. Second, the experimental results are in good agreement with the prediction (ii), and in disagreement with the theory curve (iii) which is based on approximating the homodyne output current $I(t)$ to only first order in $\left[\varphi(t)-\varphi_f(t)\right]$ so that $\bar{R}_{\rm  sq}=e^{-2r_m}$. 
Third, at the higher squeezing level the MSE is saturated, indicating the existence of an optimal squeezing level. 
Even in the theoretical curve (iv) for pure squeezed beams and zero loss, the MSE has a minimum. This curve corresponds to the fundamental limit imposed by Heisenberg's uncertainty principle for the phase and amplitude quadatures, namely $e^{-2r_p} \times e^{2r_m} \geq 1$. 
Although more squeezing decreases the $e^{-2r_m}$ term in $\bar{R}_{\rm  sq}$, it increases the $e^{2r_p}$ term due to the tracking being imperfect, which itself is a consequence of the noise in the photocurrent  (\ref{eq:Iapp and RsqSS}). This defines (self-consistently) the optimal degree of squeezing, which depends upon the parameters $|\alpha|$, $\kappa$ and $\lambda$ \cite{Sup}. 

Experimentally, we varied the amplitude $|\alpha|$, while fixing the pump beam power to 80 mW, giving squeezing and anti-squeezing levels of $-$3.2$\pm$0.2 dB and 4.9$\pm$0.3 dB, respectively. 
Theoretically, the optimal squeezing level increases with $|\alpha|$, and so too does the squeezing-enhancement, without limit. However, for our experimental conditions ($10^6 s^{-1} \leq |\alpha|^2 < 10^7 s^{-1}$) the effect of keeping the squeezing fixed is minor  (less than $3\%$ difference to $\sigma^2_s$). 
Figure 4 shows the dependence of the MSE $\sigma^2_s$ on $|\alpha|$. 
The theoretical curves show good agreement with experiments. 
 Over the whole amplitude range, the estimates with the squeezed beams surpass what is possible with coherent states, 
 with $\sigma_s^2$, averaged over the four different amplitudes, being $(15\pm4)\%$ below the CSL. 
The conclusion is essentially unaltered if one calculates the CSL not in terms of $|\alpha|^2$, but in terms of the effective photon flux $\mathcal{N}_{\rm eff}$, which equals $|\alpha|^2$ plus the extra photons due to the squeezed vacuum fluctuations in the relevant spectral range \cite{Sup}. 

We have tracked the phase of a squeezed optical field that varies stochastically in time over a significant angular range. 
Our use of Kalman filtering in real-time adaptive measurements of nonclassical systems could be applied also in solid-state and nanomechanical devices. 
Optimizing both the degree of squeezing and its bandwidth according to the experimental conditions would allow a completely rigorous treatment of photon flux. Lower losses and more squeezing would then enable a dramatic improvement to a precision that scales differently with photon flux, with $\sigma^2 \propto \mathcal{N}^{-5/8}$ \cite{Berry06} as opposed to the $\sigma^2 \propto \mathcal{N}^{-1/2}$ in the current setup.

{\noindent}{\bf Acknowledgments: }
This work was partly supported by PDIS, GIA, G-COE, APSA, FIRST commissioned by the MEXT of Japan, SCOPE program of the MIC of Japan, the Australian Research Council projects CE110001029, DP1094650 and FT100100761.

\clearpage

\noindent {\bf Fig.~1} Experimental setup. 
({\bf A}) Schematic of the experiment. 
({\bf B}) Homodyne output current versus relative phase between LO and phase-squeezed beam, and phasor diagram for a slightly non-optimal relative phase (as will occur in the phase tracking problem).
({\bf C}) Detail of the experimental setup. The abbreviations are TiS: Titanium Sapphire, CW: continuous-wave, PM: phase modulator, SHG: second-harmonic-generation, OPO: optical parametric oscillator.

\noindent {\bf Fig.~2} Time domain results of phase estimate. 
({\bf A}) Signal phase to be estimated $\varphi(t)$. ({\bf B}) Homodyne output current $I(t)$. ({\bf C}) Filtered estimate $\varphi_f(t)$. ({\bf D}) Smoothed estimate $\varphi_s(t)$. 

\noindent {\bf Fig.~3} Smoothed MSE $\sigma_s^2$ versus squeezing level. Red crosses represent experimental data. Trace (i) is the coherent-state limit which is reachable with a coherent beam only if we have unit-detection efficiency $\eta=1$. Trace (ii) is the theoretical curve from Eq.~\ref{eq:MSESM}. Trace (iii) is the theoretical curve based on a approximating the homodyne output current $I(t)$ to only first order in $\left[\varphi(t)-\varphi_f(t)\right]$ so that $\bar{R}_{\rm  sq}=e^{-2r_m}$. Trace (iv) is the theoretical curve from Eq.~\ref{eq:MSESM} for pure squeezed beams (i.e.~without loss).

\noindent {\bf Fig.~4} Dependence of the smoothed MSE $\sigma_s^2$ on the amplitude squared $|\alpha|^2$. 
Blue and red crosses are experimental data for coherent and squeezed beams, respectively. Trace (i) is the coherent-state limit. Trace (ii) is the theoretical curve for coherent beams with the experimental setup (i.e.~including inefficiency). Trace (iii) is the theoretical curve for squeezed beams, including inefficiency. Trace (iv) is the theoretical curve for pure squeezed beams and 100\% efficiency, with the squeezing level  optimized for each $|\alpha|^2$. 

\begin{figure}
  \begin{center}
   \includegraphics[width=120mm]{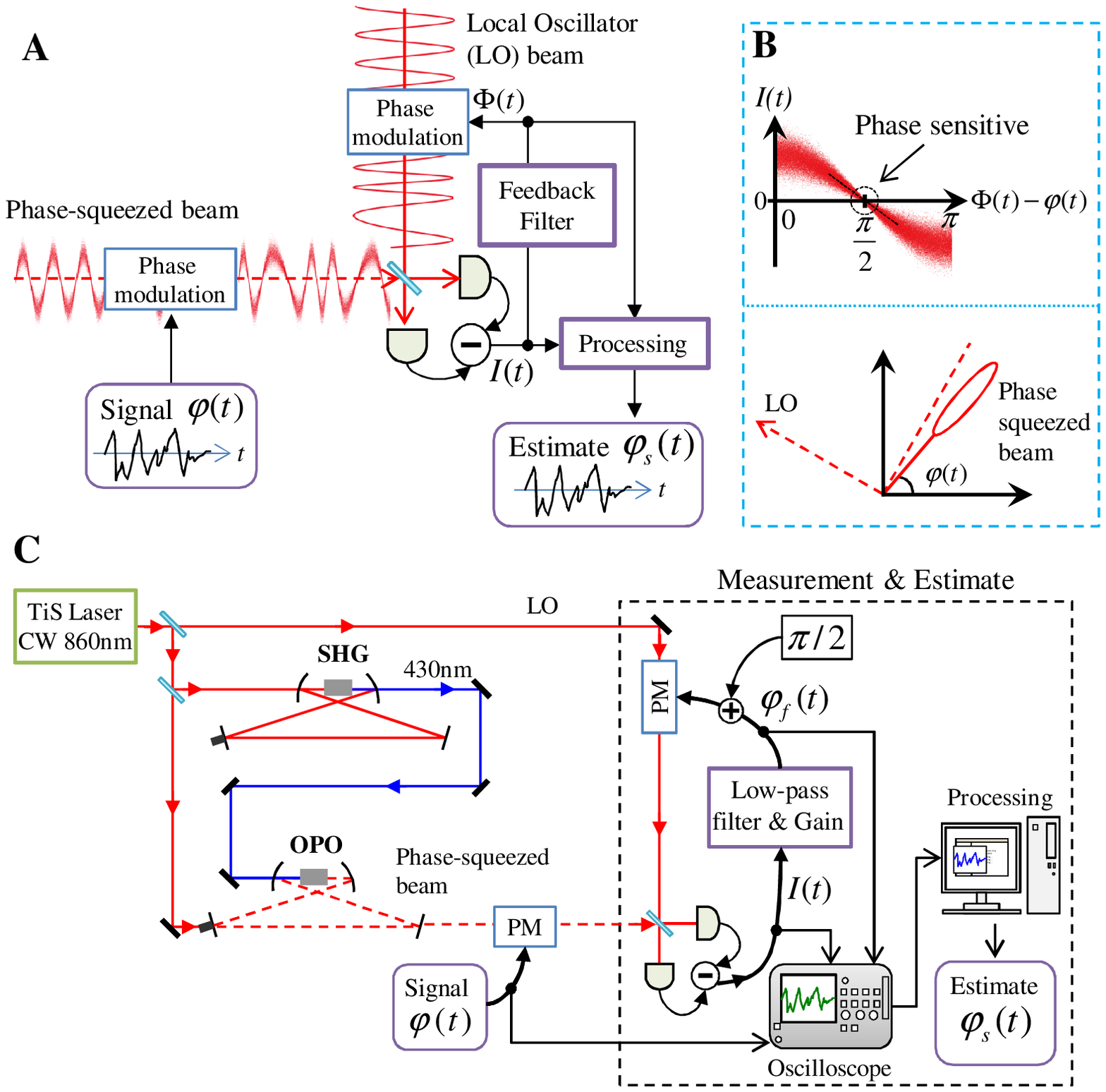}
  \end{center}
\caption{}
\end{figure}

\begin{figure}
  \begin{center}
   \includegraphics[width=80mm]{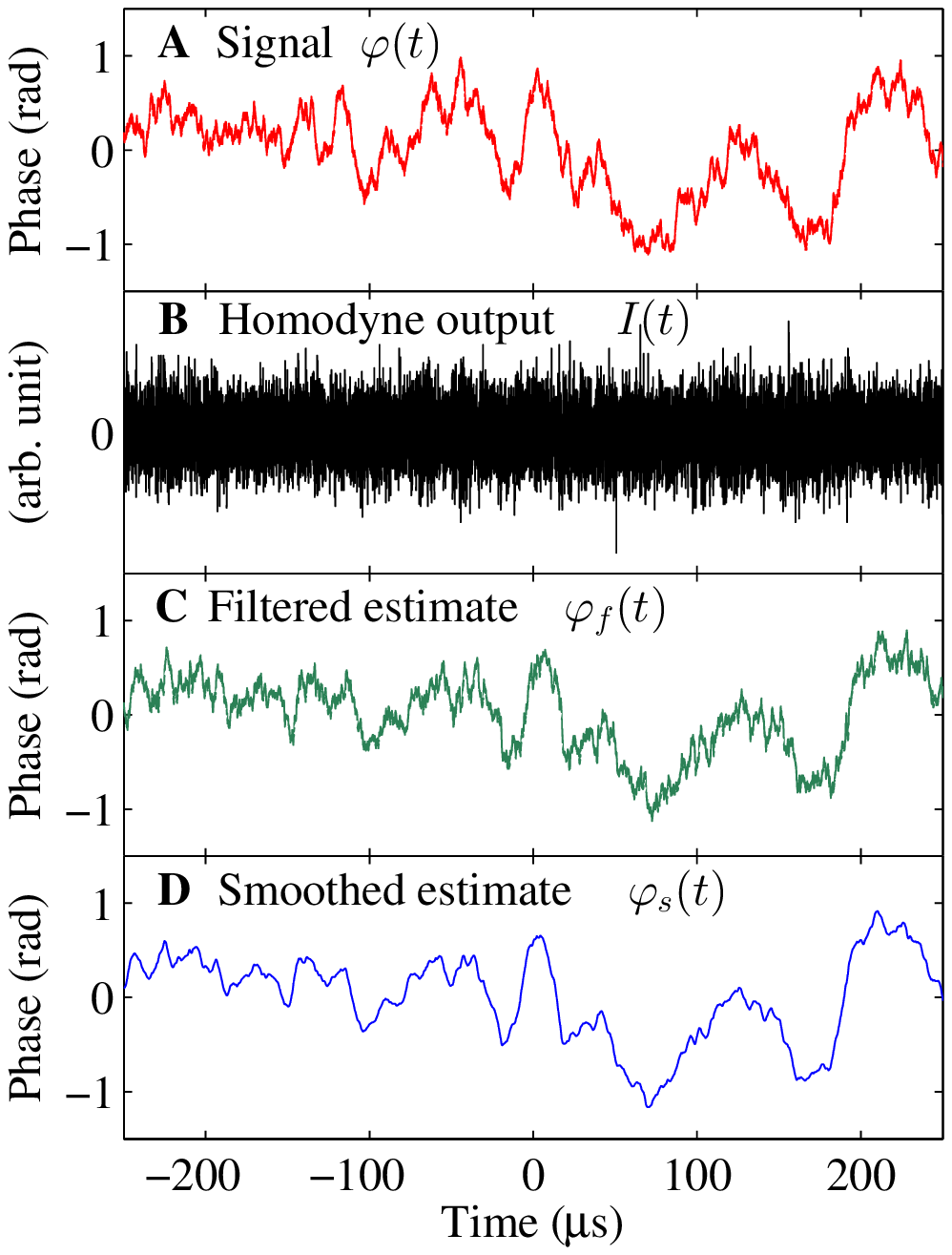}

  \end{center}
\caption{ }
\end{figure}

\begin{figure}
  \begin{center}
   \includegraphics[width=80mm]{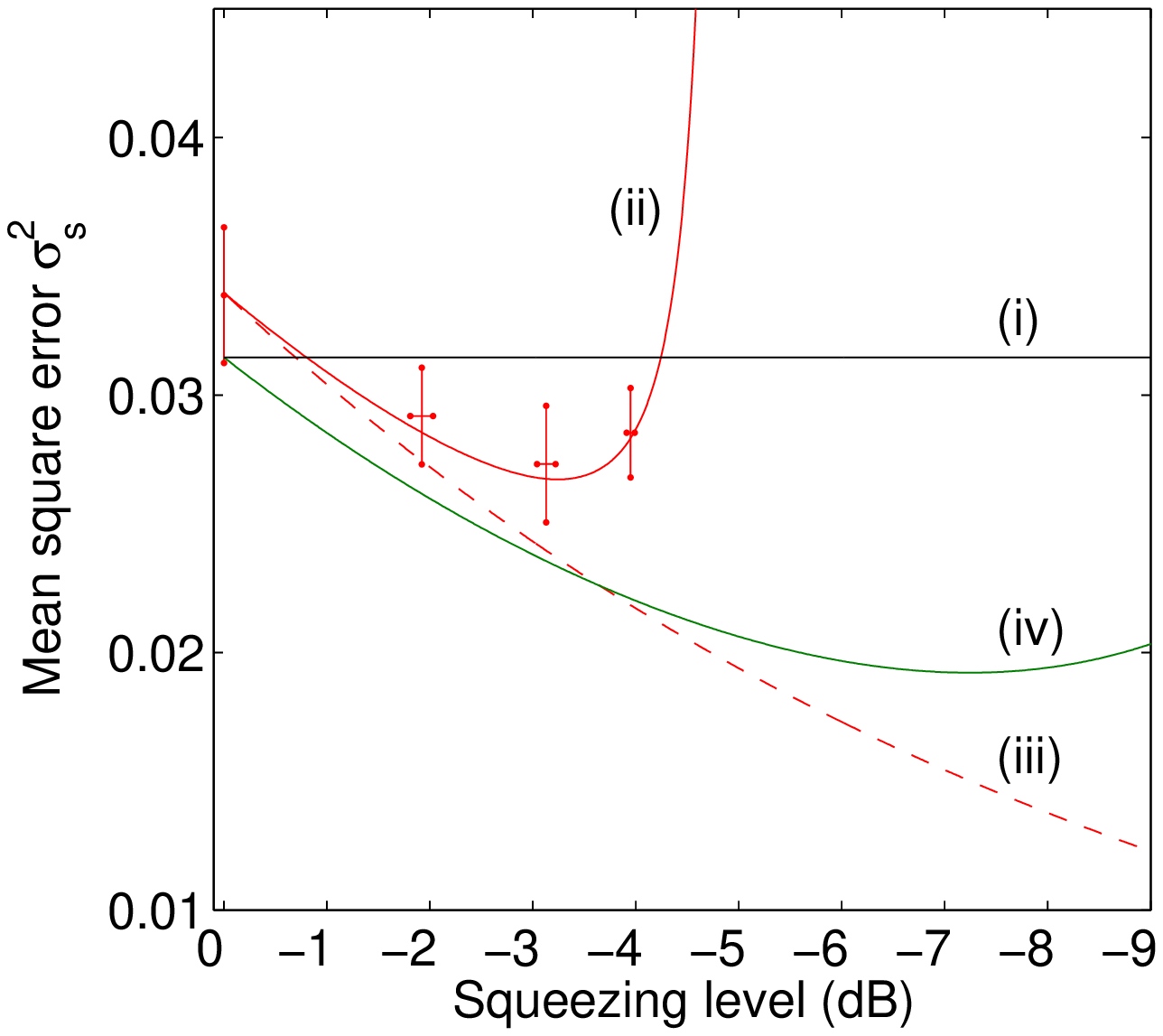}
  \end{center}
\caption{}
\end{figure}

\begin{figure}
  \begin{center}
   \includegraphics[width=80mm]{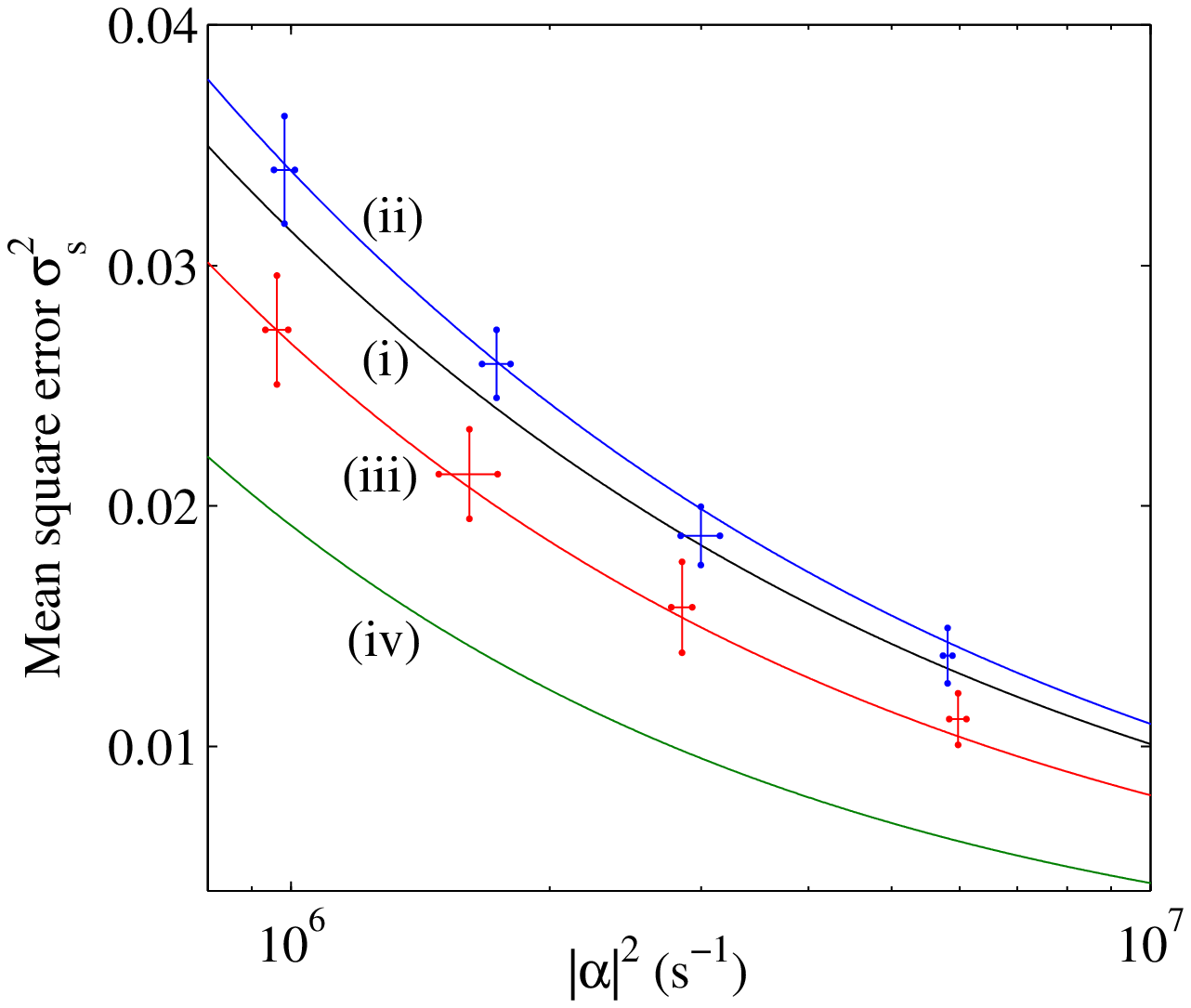}
  \end{center}
\caption{}
\end{figure}


\clearpage

\addtocounter{figure}{-4}
\renewcommand{\thefigure}{S\arabic{figure}}
\makeatletter

\begin{center}
{\LARGE
Supplementary Materials for \\
Quantum-enhanced optical phase tracking}
\end{center}

\baselineskip24pt

\section*{S1 \ Introduction}

Specific details of a more technical nature omitted in the main text are briefly discussed here. Additionally, the following sections cover a more technical discussion the experiment and details of the mathematics supporting the arguments in the main text.

Optical phase estimation has been extensively investigated because of its great importance in both science and engineering.  
As stated in the main text, phase estimation can be divided into phase sensing, where the phase is known to always lie within some small interval, and unconstrained or general phase estimation, where it is not so constrained (e.g. \cite{Xiang11}).
In phase sensing, if (as is typically the case) the field has a large coherent amplitude, the problem can be linearized in terms of the unknown phase-rotation \cite{Caves81}.
This means that the unknown phase-rotation can be treated as an unknown quadrature-displacement, considerably simplifying the problem.
The problem that we address in our paper is one of general phase estimation, where the phase cannot be approximated by a fixed quadrature (see Sec.~S3).
Moreover we consider a phase which varies continuously and stochastically in time. 

To attack this problem we use an adaptive estimation technique, and sophisticated data processing. We estimate the phase at any time with a mean square error $(15\pm4)\%$ below the coherent-state limit (CSL) --- the best precision that can be achieved with classical (i.e.~coherent) states of light. This is possible because we use nonclassical phase-squeezed light, so called because the phase uncertainty is squeezed below the CSL \cite{Caves81,Goda08,LIGO2011,Berry02,Berry06}. Because of the Heisenberg uncertainty principle, phase squeezing necessarily results in anti-squeezing in the amplitude. This anti-squeezing plays a key role in our experiment, causing there to be an optimal degree of squeezing, unlike in gravitational wave detection where more squeezing is better \cite{Caves81,Goda08,LIGO2011}. Our observation of this optimal squeezing is a new phenomenon in quantum phase estimation. 

\section*{S2 \ Details of the experiment}

Figure~S1 shows the experimental setup in more detail than shown in Fig.~1C in the main text. Specifically we show an optical mode cleaner, acousto-optic modulators (AOMs) and a radio frequency (RF) source, all of which were omitted from Fig.~1C in the main text in the interests of clarity.

The signal $\varphi(t)$, which is the waveform to be estimated, is imposed upon a weak phase-squeezed state using an electro-optic modulator. Adaptive homodyne detection of the phase-modulated beam is used to form the phase estimate.

In this section we describe the generation of the weak phase-squeezed state, and then describe its characterization using static homodyne detection. We then describe the generation of the signal and the phase estimation procedure using adaptive homodyne detection.

\begin{figure}
  \begin{center}
   \includegraphics[width=110mm]{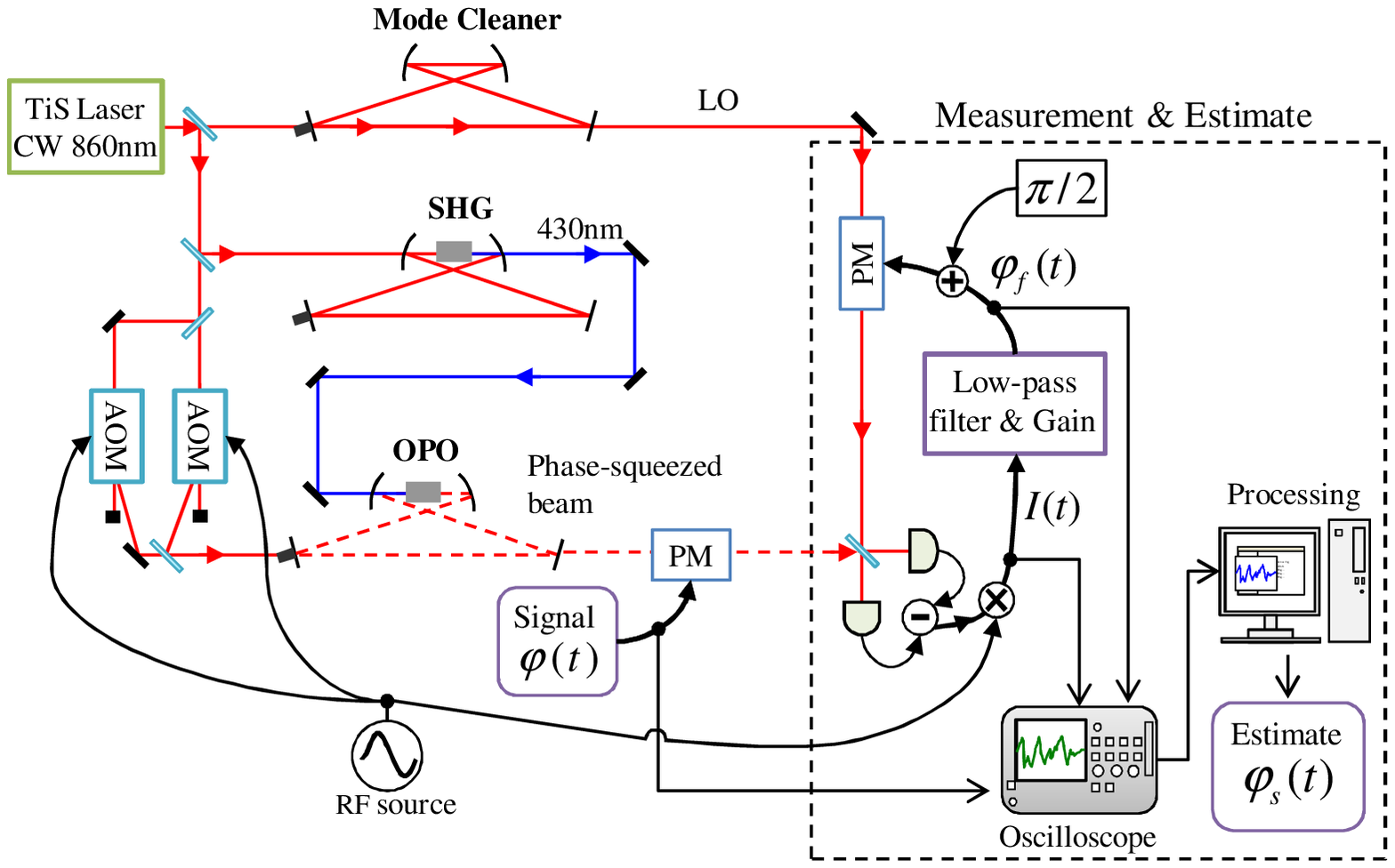}
  \end{center}
\caption{Experimental setup. The abbreviations are TiS: Titanium Sapphire, CW: continuous-wave, AOM: acousto-optic modulator, PM: phase modulator, SHG: second-harmonic-generation, OPO: optical parametric oscillator, RF source: radio frequency (5 MHz) source.
}
\end{figure}

\subsection*{S2.1 Generation and measurement of phase-squeezed states}

We use an 860 nm continuous-wave (CW) Titanium Sapphire laser in the experiment.
A fraction of the laser beam is used as the local oscillator (LO) for homodyne detection, with a spatial-mode cleaner used to improve mode-matching of the LO to the phase-squeezed state.
The phase-squeezed state is generated by a sub-threshold optical parametric oscillator (OPO) using a periodically poled KTiOPO$_4$ crystal \cite{Takeno07}.
The OPO is pumped by the 430 nm output of the second-harmonic-generator (SHG) cavity. The AOMs are used to generate a pair of optical sidebands at $\pm5$ MHz \cite{Wheatley10}, which are equivalent to a weak coherent state. The optical sidebands, which are within the 13~MHz half-width-half-maximum bandwidth of the OPO, are then injected into the OPO.

The relative phase between squeezing and the 5~MHz sidebands must be locked to ensure that a phase-squeezed beam is generated. For that purpose, we tap off 1\% of the OPO output beam to be used as a signal to lock the relative phase between the squeezing and the sidebands. Since the OPO output power is extremely low (on the order of pW), we use another homodyne detector to measure the tapped beam (not shown in Fig.~S1). The method is similar to that used in Ref. \cite{Takeno07}.

In addition to the deliberately-imposed phase modulation signal $\varphi (t)$, the experiment is subject to low-frequency phase disturbances from the environment. We use a low-frequency, low-gain feedback loop to suppress the environmental disturbances. When we use static homodyne detection to characterize the squeezing (anti-squeezing) level of the phase-squeezed states before performing the phase estimation experiment, we use only this classical feedback loop to lock the relative phase between the LO and the phase-squeezed beams.

With a LO beam power of 2 mW, the shot noise level of the homodyne detector is 10.4 dB higher than the circuit noise level (that is, the ratio of the measured shot noise to circuit noise is $S=11.0$ ). The homodyne visibility is $\xi=0.988$, the optical transmission is $\rho=0.97$, and the quantum efficiency of each of the photo-diodes is $\zeta=0.99$.
The overall efficiency of the homodyne detection is thus calculated as $\eta= \xi^2\rho\zeta (S-1) /S = 0.85$ \cite{Appel07}.

Figure S2 shows static homodyne measurements of the phase-squeezed state when the power of the OPO pump beam is 180~mW. Fig.~S2 shows the noise spectrum as measured with a radio frequency spectrum analyzer in the region around the 5~MHz sidebands. As we have mentioned previously, these are static homodyne measurements of the phase-squeezed state. So for Fig.~S2 we do not apply phase modulation signal $\varphi (t)$ to the phase-squeezed beam, and do not use the phase-tracking loop. The residual peak at around 5 MHz in trace (ii) comes from fluctuations of the classical locking loop. Compared to our signal bandwidth ($\lambda/2\pi = 9.4$ kHz), the bandwidth of the squeezing is broad enough that we may assume that the squeezing is broadband over the frequency range of interest.

In our experiment, we use pump beam powers of 0, 30, 80 and 180~mW. Squeezing and anti-squeezing levels around 5 MHz are shown in Fig.~S3. The solid line is a theoretical curve fitted to the experimental results with the overall loss $l_{\rm {} sq}=1-\eta(1-l_{\rm{}s})$ (comprising the homodyne detection efficiency, $\eta$, and the loss from the OPO and phase modulator, $l_{\rm{}s}$) being a free parameter.
We define the measured squeezing and anti-squeezing levels as $R_{-}:=e^{-2r_{m}}$ and $R_{+}:=e^{2r_{p}}$. With an overall loss $l_{\rm {} sq}$ and a pure squeezing parameter $r$, the measured squeezing and anti-squeezing levels can be written as \cite{LeonhardtEQO},
      \begin{eqnarray}
         R_{-} &:=& e^{-2r_m} = (1-l_{\rm {} sq})e^{-2r}+l_{\rm {} sq},
          \nonumber \\
         R_{+} &:=& e^{2r_p} = (1-l_{\rm {} sq})e^{2r}+l_{\rm {} sq}.
      \end{eqnarray}
Hence we obtain the following equation,
      \begin{equation}
\label{eq:antisq}
         R_{+}= \frac{(1-l_{\rm {} sq})^2}{R_{-}-l_{\rm {} sq}}+l_{\rm {} sq}.
      \end{equation}
In Fig.~S3, we obtain the overall loss $l_{\rm {} sq}=0.33$ which mainly comes from the OPO and phase modulator.

\begin{figure}[h]
 \begin{center}
  \begin{tabular}{cc}
    \begin{minipage}{0.47\hsize}
      \begin{center}
       \includegraphics[width=60mm,clip]{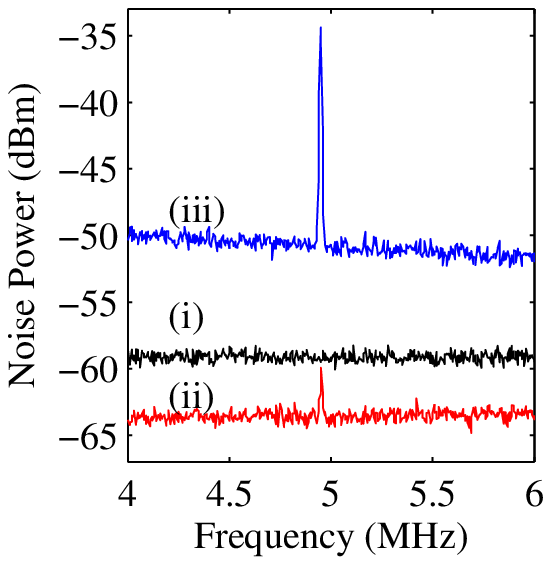}
       \caption{Characterizing the phase squeezing. The data is recorded with a spectrum analyzer. Resolution and video bandwidths are 10 kHz and 300 Hz, respectively. Trace (i) is shot noise level. Traces (ii) and (iii) are the squeezing and anti-squeezing levels where the LO phase is locked to those quadratures. }
      \end{center}
    \end{minipage}
&
    \begin{minipage}{0.47\hsize}
      \begin{center}
       \includegraphics[width=60mm,clip]{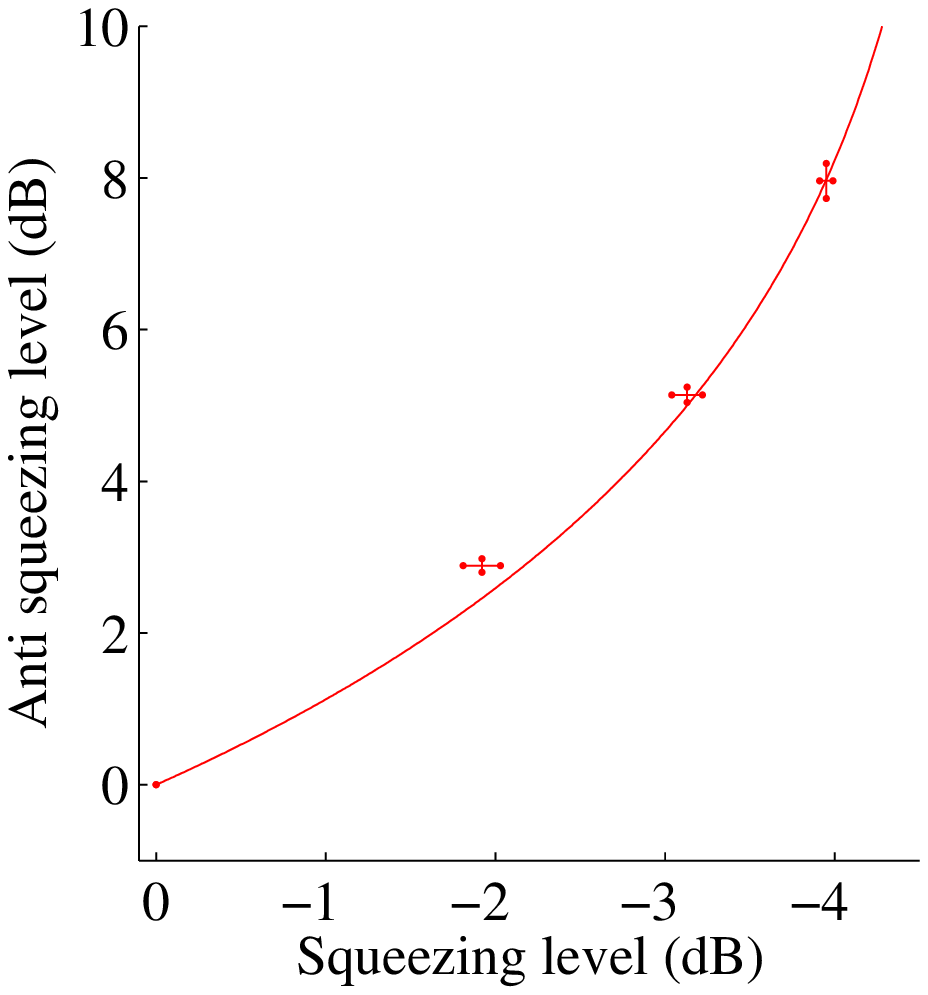}
       \caption{Anti-squeezing level versus squeezing level. Solid curve is fitted to the experimental data with loss being a free parameter. The overall loss $l_{\rm {} sq}$ is obtained as 0.33.
}
      \end{center}
    \end{minipage}
  \end{tabular}
 \end{center}
\end{figure}

\subsection*{S2.2 Phase Tracking}

The signal waveform to be estimated is generated by a random signal source and a low-pass filter with a cutoff frequency of $\lambda/2\pi = 9.4$ kHz. The signal source is a digital signal generator (AFG3021, Tektronix), which generates white Gaussian noise (3dB bandwidth is 25 MHz).
The signal $\varphi(t)$ is imposed upon the phase-squeezed beam using an electro-optic modulator (LM0202, Linos). We perform homodyne detection on this phase-modulated beam.
After homodyne detection of the phase-modulated beam, the homodyne current is demodulated at 5 MHz and then goes to the tracking filter (another low-pass filter with a cutoff frequency $\lambda / 2\pi$ according to the Kalman filter theory).
The output of the tracking filter gives the filtered estimate $\varphi_f(t)$, which is then shifted by $\pi$/2 and applied to the phase of the LO beam with another electro-optic modulator (waveguide modulator, EOSPACE ). The gain of the phase-lock loop is optimized to give the maximum sensitivity.
As explained previously, an additional low-frequency, low gain feedback loop is used to isolate the experiment from environmental disturbances.

We record $\varphi(t)$, $\varphi_f(t)$ and the demodulated homodyne current $I(t)$ using an oscilloscope with a sampling rate of 100~MHz.
We calculate the mean square error from 2 ms of data, comprising $2\times10^5$ data points.
To calculate the average and the standard deviation of the mean square error, we repeat the measurement 15 times.
Note that each the correlation time of the noise signal $\varphi (t)$ is approximately $\lambda^{-1}$, and so $2\times10^5$ data points have effectively $2 {\rm \ ms} / \lambda^{-1} \sim 10^2$ independent data points.

\section*{S3 \ Phase estimation with broadband squeezing}
In this section we describe phase estimation with broadband squeezing.
As in the main text, the phase variation $\varphi(t)$ obeys the following,
      \begin{equation}
         \varphi(t)=\sqrt{\kappa} \int^t_{-\infty} e^{-\lambda(t-s)} dV(s),
       \label{eq:OUnoise}
      \end{equation}
where $dV(s)$ is a Wiener increment which satisfies $\left< dV(s) dV(s') \right>=\delta(s-s')(dt)^2$, $\lambda$ is the bandwidth of the phase noise $\varphi$, and $\kappa/2\lambda$ is the mean square variation of $\varphi$. Note that in our experiment this variation is much larger than mean square error with which we can estimate $\varphi$.
The phase-modulated beam is measured by homodyne detection. The LO phase $\Phi(t)$ is adaptively controlled to be $\Phi(t) := \varphi_f(t) + \pi/2$, where $\varphi_f(t) \approx \varphi(t)$ is our filtered estimate.
In our experiment the squeezing bandwidth is many MHz, which is much larger than the bandwidth $\lambda$ of the phase noise. Hence we can approximate the squeezing noise as white.
 Using the normalization of \cite{Berry02}, the homodyne output current is
      \begin{eqnarray}
         I(t)dt &=& 2\left| \alpha \right |
                      \sin \left[
                                   \varphi(t)-\varphi_f(t)
                      \right] dt
                 +\sqrt{R_{\rm {} sq}(t)} dW(t),
       \label{eq:It} \\
         R_{\rm {} sq}(t) &=&
                       \sin^2 \left[ \varphi(t)-\varphi_f(t) \right] e^{2r_p}
                     + \cos^2 \left[ \varphi(t)-\varphi_f(t) \right] e^{-2r_m},
       \label{eq:Rsqt}
      \end{eqnarray}
where $|\alpha|^2$ is the photon flux of the coherent component of the beam (i.e. photons per unit time), $dW(t)$ is a Wiener increment representing quantum white noise, satisfying $\left< dW(s) dW(s') \right>=\delta(s-s')(dt)^2$,
and the amplitude of this noise $R_{\rm sq}$ is determined by the squeezing
($r_m \geq 0$) and anti-squeezing ($r_p \geq 0$) parameters defined earlier.
The squeezing factor $R_{\rm sq}(t)$ is time-dependent because of
tracking error $\varphi(t)-\varphi_f(t)$, the unknown and varying misalignment between the
squeezing ellipse and the local oscillator phase.

We denote the tracking error by $\Delta_f(t) = \varphi(t)-\varphi_f(t)$,
and its stationary ensemble average by $\sigma_f^2 = \left< \Delta_f^2 \right>_{\rm ss}$.
Because $\sigma_f \ll 1$ in our experiment we can approximate $I(t)dt$ by a 2nd order expansion in $\Delta_f(t)$.
This gives
      \begin{eqnarray}
          I(t)dt &\simeq& 2\left| \alpha \right |
                      \Delta_f(t) dt
                 +\sqrt{R_{\rm sq}} dW(t), \label{eq:Iapp1} \\
        R_{\rm sq}(t) &\simeq&
                         \Delta_f^2 (t) e^{2r_p} + [1-\Delta_f^2(t)] e^{-2r_m}.
       \label{eq:Rsq}
      \end{eqnarray}

The approximations given here help to illustrate the difference between phase estimation and phase sensing.
In phase sensing, the variation in the system phase $\varphi(t)$ is sufficiently small that $I(t)dt$ can be linearised in $\varphi(t)$.
In phase estimation, this is not the case and a better approximation is needed.
Here the filtered estimate $\varphi_f(t)$ is sufficiently good that the sine function can be linearised in $\Delta_f(t)$.
(This is why feedback is vital in phase estimation.)
Furthermore, in the case where there is squeezing, there is a difference in the treatment of the squeezing term $R_{\rm sq}(t)$.
In phase sensing, the variation in the system phase is sufficiently small that one may simply take $R_{\rm sq}(t)\simeq e^{-2r_m}$.
One can simply treat the phase-rotation as a displacement of the squeezed quadrature, and not consider the anti-squeezing.
In contrast, for phase estimation (even with feedback) the variation of the system phase is sufficiently large that one needs a better approximation for $R_{\rm sq}(t)$.
Because the filtered estimate is not perfectly accurate, there is a significant contribution to the error from the anti-squeezed quadrature.

A further approximation is to replace $\Delta_f^2(t)$ by its stationary average $\sigma_f^2$, which is independent of time. (This approximation is justified below by comparing the time scale of its variation to that of the subsequent filter.)
The normalized homodyne output current can then be written as
      \begin{eqnarray}
          I(t)dt &\simeq& 2\left| \alpha \right |
                       \Delta_f(t)dt
                 +\sqrt{\bar R_{\rm sq}} dW(t), \label{eq:Iapp} \\
        \bar R_{\rm sq} &=&
                         \sigma_f^2 e^{2r_p} + (1-\sigma_f^2) e^{-2r_m},
       \label{eq:RsqSS}
      \end{eqnarray}
where $\bar R_{\rm sq}$ is an effective (time-independent) squeezing factor which takes into account the tracking error in an average sense. A measurement better than the coherent-state limit is expected to be possible when
this $\bar R_{\rm sq}$ is less than $1$.

For an equation of the form of Eq.~\ref{eq:Iapp}, it is known that the optimal causal estimate (as required for tracking) is provided by the Kalman filter \cite{Tsang09PRA}:
      \begin{equation}
          \varphi_f(t) = \Gamma \int^t_{-\infty} e^{-\lambda(t-s)}
                                    \frac{I(s)}{2|\alpha|} ds,
        \label{eq:feedback}
      \end{equation}
where $\Gamma$ is the Kalman gain.
Before going further, we return to the approximation in Eq.~\ref{eq:RsqSS}, where we replace the time-varying $\Delta_f^2(t)$, with its average $\sigma_f^2$.
The characteristic time of the Kalman filter is $\lambda^{-1}$; that is, the homodyne output current $I(t)$ is, roughly speaking, averaged over the period of $\lambda^{-1}$. Thus if the variation of $\Delta_f^2(t)$ is rapid compared to $\lambda^{-1}$, it is justified to replace the time-varying $\Delta_f^2(t)$ by the average $\sigma_f^2$.
Figure S4 shows a typical experimental trace of the temporal variation of the squared filtered estimate error $\Delta_f^2(t) = [\varphi(t)-\varphi_f(t)]^2$, and also its auto-correlation function.
The experimental conditions are the same as Fig.~2 in the main text.
The variation of the squared filtered estimate error has a correlation time around 2$\mu$s. The Kalman filter characteristic time is calculated as $\lambda^{-1}\simeq 17 \mu s$ which is much larger than the correlation time of the squared filtered estimate error. Therefore our approximation in Eq.~\ref{eq:RsqSS} turns out to be appropriate.

\begin{figure}[ht]
      \begin{center}
       \includegraphics[width=100mm,clip]{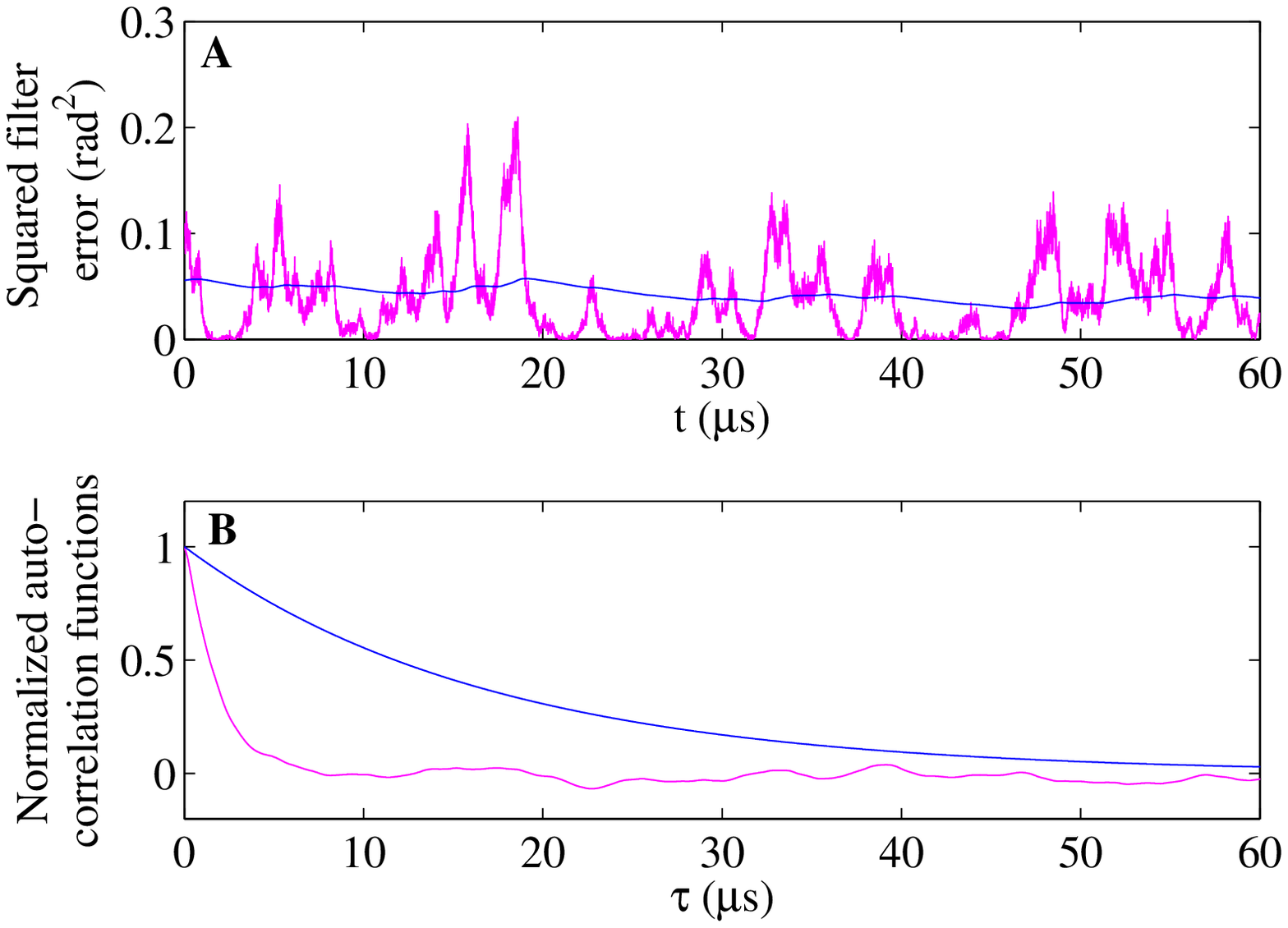}
       \caption{({\bf A}) Squared error of the filtered estimate $\Delta_f^2$ under the same conditions as Fig.~2 in the main text. The purple trace is $\Delta_f^2$, and the blue trace is that after the Kalman filter.
({\bf B}) Normalized auto-correlation $\left \langle \left[\Delta_f(t)\right]^2,\left[\Delta_f(t+\tau)\right]^2 \right \rangle$ as measured in the experiment (purple curve), and theoretical curve for the Kalman filter of white noise (blue curve).}
      \end{center}
\end{figure}

This separation of time scales is a consequence of the fact that our experiment was performed in the limit $\Gamma^{-1} \ll  \lambda^{-1}$ (for the traces in Fig.~S4, $\Gamma^{-1} \approx 4 \mu s$). This can be seen by considering the differential equation for $\Delta_f(t)$. First, combining Eqs. \ref{eq:Iapp} and \ref{eq:feedback} gives
\begin{equation}
          \varphi_f(t) = \Gamma \int^t_{-\infty} e^{-\lambda(t-s)}
                        \left [
                              \Delta_f(s) ds
                             +\frac{\sqrt{ \bar R_{\rm sq} } }
                                   {2 \left | \alpha \right | } dW(s)
                       \right ].
          \label{eq:feedback2}
      \end{equation}
Combining this with Eq.~\ref{eq:OUnoise} gives
\begin{eqnarray}
   d\Delta_f &=& d\varphi(t)-d\varphi_f(t) \\
             &=& -(\lambda + \Gamma) \Delta_f(t)dt
                + \sqrt{\kappa} dV(t)
                - \frac{\Gamma \sqrt{ \bar R_{\rm sq} } }
                {2 \left | \alpha \right | } dW(t).
    \label{SDEforDelta}
   \end{eqnarray}
Thus, under the above approximation, $\Delta_f$ has a correlation time of $(\lambda+\Gamma)^{-1}$, and $\Delta_f^2$ a correlation time of half that.
For $\Gamma \gg \lambda$, the correlation time for $\Delta_f^2$ is much less than that of the filter, and hence on the time scale of the filter $\Delta_f^2$ may be replaced by its ensemble average. More fundamentally, this difference in time scales comes because we have conducted our experiment in the regime where $\epsilon \equiv\sqrt{\lambda^2 \bar{R}_{\rm sq}/ (4 \left|\alpha\right|^2 {\kappa})}$ is small, as in this regime $\Gamma \simeq \lambda / \epsilon$ (see below).

Even if we do not replace $R_{\rm sq}$ (\ref{eq:Rsq}) by $\bar{R}_{\rm sq}$ (\ref{eq:RsqSS}), we find the same result for $\sigma_f^2$. This follows from the differential equation for $\an{\Delta_f^2}$ using Eq.~\ref{eq:Rsq}, namely
   \begin{equation}
     d\an{\Delta_f^2} =
                   -2 (\lambda + \Gamma) \an{\Delta_f^2} dt
                    + \kappa dt- \frac{\Gamma^2
                      \left< \Delta_f^2 e^{2r_p} + (1-\Delta_f^2) e^{-2r_m}
                      \right>         }
                      {4 \left | \alpha \right |^2 } dt.
   \end{equation}
This gives a linear equation for $\sigma_f^2 = \langle \Delta_f^2 \rangle_{\rm ss}$ which is identical to the equation which would be obtained from Eq.~\ref{SDEforDelta} in which $\bar{R}_{\rm sq}$ contains
$\sigma_f^2$ already. This means that our theory models the experiment for any value of $\epsilon$ (provided
$\sigma_f^2$ remains small), but our argument for the optimality of the Kalman filter, and the corresponding smoothed estimator,
holds only for $\epsilon$ small. This is reasonable for our experiment, in which $ \epsilon \leq 0.2 $.

Returning to Eq.~\ref{eq:feedback2}, it is obvious that phase estimation using the phase-squeezed beam with the amplitude $|\alpha|$ and squeezing factor $\bar R_{\rm sq}$ is mathematically equivalent to that using a coherent beam with the amplitude $|\alpha|/\sqrt{\bar R_{\rm sq} }$. (Note that this equivalence holds in terms of the estimate precision, not in terms of photon number.)
Therefore we can obtain the Kalman gain $\Gamma$ and the mean square error $\sigma_f^2$ by formally replacing $|\alpha|$ in Ref.~\cite{Tsang09PRA} with $|\alpha|/\sqrt{\bar R_{\rm sq}}$, giving
      \begin{eqnarray}
         \Gamma &=& -\lambda
               + \sqrt{
                    \lambda^2+4\kappa \left| \alpha \right|^2/\bar R_{\rm sq}
                      },
               \\
              \sigma_f^2 &=& \frac{\lambda \bar R_{\rm sq}}{4\left| \alpha \right|^2}
                        \left(
                          \sqrt{1+
                                  \frac{4\kappa\left| \alpha \right|^2}
                                       {\lambda^2\bar R_{\rm sq}}
                               }
                           -1
                        \right).
      \end{eqnarray}

In the limit $\epsilon \ll 1$,
the optimal gain is $\Gamma \approx \lambda/\epsilon$, and $\sigma_f^2 \approx \sqrt{\kappa \bar{R}_{\rm  sq}}/\left( 2\left| \alpha \right| \right)$.
All of these equations are still implicit because $\bar{R}_{\rm sq}$ (\ref{eq:RsqSS}) is a function of $\sigma_{f}^{2}$.
Solving the implicit equation for $\sigma_{f}^{2}$ gives explicitly, and exactly (i.e. without assuming $\epsilon \ll 1$)
      \begin{eqnarray}
        \Gamma   &=& -\lambda
               -\frac{\kappa \left( e^{2r_p}-e^{-2r_m} \right)}{2e^{-2r_m}}
               +\frac{1}{2} \sqrt{
                              \left(
                               \frac{\kappa \left( e^{2r_p}-e^{-2r_m} \right) }
                                    {e^{-2r_m}}+2\lambda
                              \right)^2
                             +\frac{16|\alpha|^2 \kappa}{e^{-2r_m}}
                                 },
               \label{eq:KalmangainG} \\
                            \sigma_f^2     &=& \frac{1}{8|\alpha|^2+4\lambda(e^{2r_p}-e^{-2r_m})}
                        \biggl[
                           -2 \lambda e^{-2r_m}+\kappa (e^{2r_p}-e^{-2r_m})
               \nonumber \\
                  & & \  +\sqrt{
                            \left[2 \lambda e^{-2r_m}
                                  +\kappa (e^{2r_p}-e^{-2r_m})
                            \right]^2
                             +16|\alpha|^2 \kappa e^{-2r_m}
                               }
                        \ \biggr].
             \label{eq:sigmafBB}
\end{eqnarray}

 As in Ref.~\cite{Wheatley10}, we can apply the smoothing technique of Ref.~\cite{Tsang09PRA} for further improvement.
The smoothed estimate $\varphi_s(t)$ and mean square error $\sigma_s^2=\left< \left[ \varphi(t)-\varphi_s(t) \right]^2\right>$ are
      \begin{eqnarray}
           \varphi_s(t) &=& \left( 2\lambda+\Gamma \right)
                        \int^\infty_{t} e^{-(\lambda+\Gamma)(s-t)}
                                               \varphi_f(s) ds, \\
           \sigma_s^2 &=& \frac{\kappa}
                               {2\sqrt{
                               4\kappa \left| \alpha \right|^2/\bar R_{\rm sq}
                               +\lambda^2
                                             }}.
       \label{eq:MSES}
      \end{eqnarray}
These formulae are obtained by formally replacing $|\alpha|$ in Ref.~\cite{Tsang09PRA} with $|\alpha|/\sqrt{\bar R_{\rm sq}}$. In the limit $\epsilon \ll 1$, we obtain the simple relation $\sigma_s^2 \approx \sigma_f^2/2$
as in Ref.~\cite{Wheatley10}.

\section*{S4 \ Optimum squeezing level for phase tracking}

In this section, we discuss the optimum squeezing level for the phase tracking, as quantified by the mean square error in the smoothed estimate.

From Eq.~\ref{eq:RsqSS} and \ref{eq:MSES}, it is obvious that highly squeezed states would not improve the phase tracking if there were too much anti-squeezing noise. Quantum mechanics does not allow us to squeeze a physical quantity without anti-squeezing the canonically conjugate counterpart. Hence there must be an optimum squeezing level for the phase estimation problem.
Figure S5 shows the smoothed mean square error $\sigma_s^2$ as a function of both squeezing and anti-squeezing levels,
with other parameters ($\alpha$, $\kappa$, $\lambda$) fixed.
In this figure the lower-right half is forbidden due to the uncertainty principle.
Even for a pure squeezed state (at the boundary of the forbidden region), the minimum mean square error is obtained for an optimum squeezing level of about 7 dB.
In practice, with finite loss, the squeezed state is not pure, and it is not possible to attain the boundary with the forbidden region.
The minimum anti-squeezing for a given squeezing will follow a curve away from the boundary, according to Eq.~\ref{eq:antisq}.
The curve for the loss in our experiment is shown in Figure S5.
The qualitative behavior is the same as in the pure state case.
There is an optimal squeezing level, and increasing the squeezing beyond this will increase the mean square error.

\begin{figure}[ht]
      \begin{center}
       \includegraphics[width=110mm,clip]{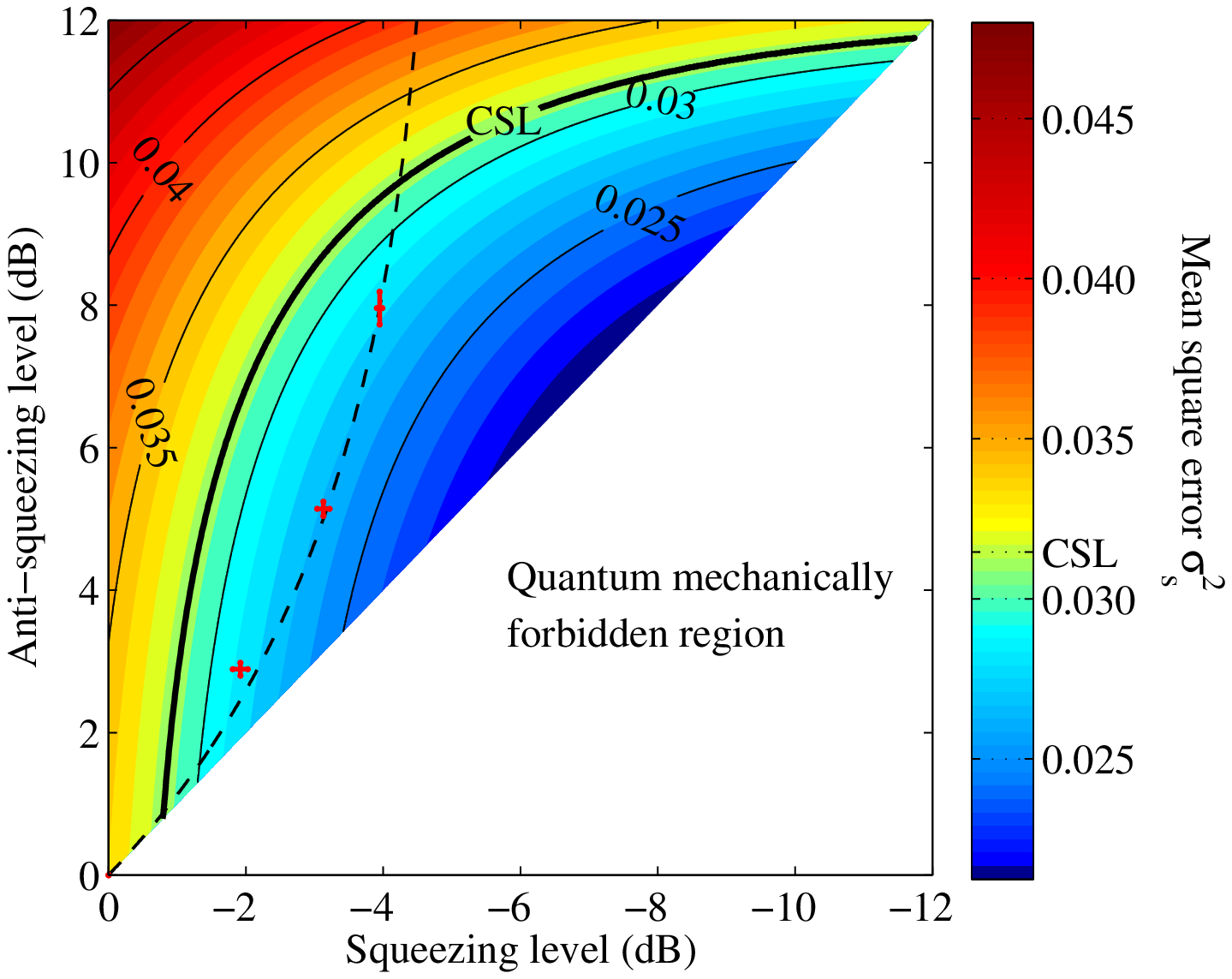}
       \caption{ Mean square error of the smoothed estimate, as a function of both squeezing and anti-squeezing levels, calculated from Eq.~\ref{eq:MSES}. Other parameters ($\alpha$, $\kappa$, $\lambda$) are fixed and the same as Fig.~3 in the main text. The black dashed line and red crosses are the predicted and experimentally observed squeezing and anti-squeezing levels in our setup (see Fig.~S3). CSL stands for the coherent-state limit.
}
      \end{center}
\end{figure}

The optimum squeezing level varies with parameters $\alpha$, $\kappa$ and $\lambda$. From Eq.~\ref{eq:RsqSS}, it would be obvious that the optimum squeezing level is higher if the mean square error of the filtered estimate $\sigma_f^2$ is smaller. This is because the anti-squeezing noise has less effect if $\sigma_f^2$ is small.
The mean square error $\sigma_f^2$ would be reduced by increasing $\alpha$, or $\lambda$, or by reducing $\kappa$. Note that the mean square variation of signal $\varphi$ is $\kappa/2\lambda$, and the estimate error $\sigma_f^2$ is smaller with smaller signal variation (smaller $\kappa$ or larger $\lambda$). 
Therefore, the optimum squeezing level is higher if $\alpha$ or $\lambda$ is larger or $\kappa$ is smaller (see Fig. S6).

\begin{figure}[ht]
      \begin{center}
       \includegraphics[width=125mm,clip]{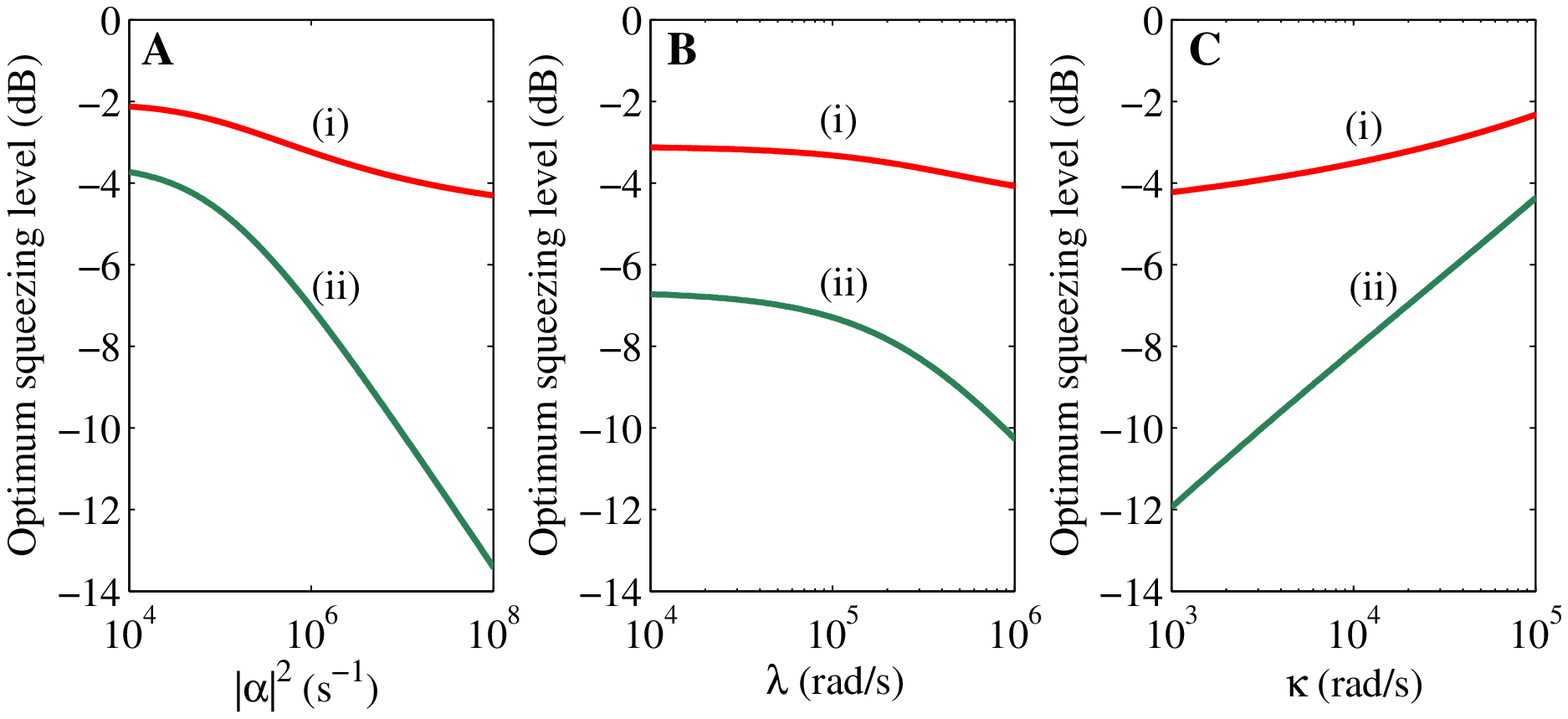}
       \caption{Optimum squeezing levels as a function of ({\bf A}) $|\alpha|^2$, ({\bf B}) $\lambda$ and ({\bf C}) $\kappa$. The optimum squeezing levels are calculated by minimizing $\sigma_s^2$ of Eq.~\ref{eq:MSES}. The parameters other than the varying parameter are fixed and the same as Fig.~3 in the main text (e.g., in ({\bf A}), $\kappa$ and $\lambda$ are fixed and the same as Fig.~3 in the main text.). Traces (i) are for squeezed beams in our experimental setup (i.e. including loss). Traces (ii) are for pure squeezed beams.
}
      \end{center}
\end{figure}

\section*{S5 \ Squeezing bandwidth and photon flux}

In this section, we derive the photon flux due to squeezing, as this is required (see Sec.~S7) for a complete
accounting of the resources used in phase estimation. As we shall show shortly, infinite-bandwidth squeezing leads to an infinite photon flux. Therefore, finite-bandwidth squeezing must be considered explicitly.

Let us consider a squeezing spectrum $R_{-}(\Omega)$ and anti-squeezing spectrum $R_{+}(\Omega)$ of the standard form \cite{Takeno07}
      \begin{eqnarray}
        R_{\pm}(\Omega) &=&
                       1+(R_{\pm}-1)
                           \frac{\left[(1\mp x) \Delta \Omega_{0}\right ]^2}
                                {\Omega^2+
                                  \left[(1\mp x) \Delta \Omega_{0}\right ]^2},
       \label{eq:ROmega} \\
             x &=& \frac{R_{+}-R_{-}-2\sqrt{(1-R_{-})(R_{+}-1)} }
                            {R_{+}+R_{-}-2},
       \label{eq:x}
      \end{eqnarray}
where $R_{\pm}$ are squeezing and anti-squeezing levels at the center frequency ($\Omega=0$). In the case of the OPO, $2\Delta \Omega_0$ and $x$ correspond to the cavity's decay rate and the normalized pump amplitude respectively. The difference between the bandwidths for squeezing $(1+x) \Delta \Omega_{0}$ and anti-squeezing $(1-x) \Delta \Omega_{0}$ ensures that $R_{+}(\Omega) R_{-}(\Omega) =1$ for all $\Omega$ when the squeezed state is pure.

 The number of photons due to squeezing in a frequency mode $\Omega$ is given as \cite{LeonhardtEQO},
      \begin{equation}
          n(\Omega)=\frac{1}{4}
                    \left[
                           R_{+}(\Omega)+R_{-}(\Omega)
                   \right]
                   -\frac{1}{2}.
      \end{equation}
 The photon flux per unit time $\mathcal{N}_{\rm {} sq}$ due to the squeezing is given by integrating $n(\Omega)$ over $\Omega$ \cite{GardinerQN},
      \begin{equation}
        \mathcal{N}_{\rm {} sq}=\frac{1}{2\pi}
                          \int^{\infty}_{-\infty} n(\Omega) d\Omega
                        = \frac{1}{4} \left[(R_{+}-1) (1- x)
                               +(R_{-}-1) (1+ x) \right]
                           \frac{\Delta \Omega_{0}}{2}.
       \label{eq:Nsq}
      \end{equation}
Note that, as mentioned previously, the photon flux diverges for infinite-bandwidth squeezing.

\section*{S6 \ Phase estimation with { finite-bandwidth }squeezing}

In this section, we derive mean square error of the phase estimate when the squeezing bandwidth is finite.
This will be used in Sec.~S7 to calculate the (small) effect that restricting the bandwidth of our squeezing
would have had.

 For finite-bandwidth squeezing the equations \ref{eq:Iapp} and \ref{eq:RsqSS} for the homodyne output current must be
modified to
      \begin{eqnarray}
       I(t)dt &\simeq& 2\left| \alpha \right |
                  \left[
                             \varphi(t)-{ \tilde\varphi_{f} }(t)
                 \right] dt
                +dY(t), \label{eq:IappN} \\
        \left \langle dY(s) dY(s')\right \rangle
            &=&  dsds' \left[
               { \tilde\sigma_{f} }^2 X_{+}(s-s') + (1-{ \tilde\sigma_{f} }^2) X_{-}(s-s')
                      \right],
       \label{eq:dYAutoCF}
      \end{eqnarray}
where $X_{\pm}(s-s')$ are auto-correlation functions for the squeezed ($-$) and anti-squeezed ($+$) quadratures, and the tilde is used to distinguish particular finite-bandwidth properties from the previously calculated infinite-bandwidth quantities (with no tilde).
$X_{\pm}(s-s')$ are calculated by the inverse Fourier transform of the squeezing and anti-squeezing spectrum $R_{\pm}(\Omega)$,
      \begin{eqnarray}
       X_{\pm}(s-s') &=& \frac{1}{2\pi}
                  \int^{\infty}_{-\infty}
                     R_{\pm}(\Omega) e^{i\Omega (s-s')}d\Omega
          \nonumber \\
                    &=& \delta(s-s')
                    +(R_{\pm}-1)
                      \frac{(1\mp x)\Delta\Omega_{0}}{2}
                       e^{-(1\mp x)\Delta \Omega_{0} |s-s'|}.
       \label{eq:ACFXpm}
      \end{eqnarray}

From Eqs. \ref{eq:feedback} and \ref{eq:IappN}, and assuming a filter of the same form as above,
the estimator used for tracking is

      \begin{equation}
        { \tilde\varphi_{f} }(t) = { \tilde\Gamma} \int^t_{-\infty}
                            e^{-\lambda(t-s)} \frac{I(s)}{2|\alpha|} ds
                          = { \tilde\Gamma} \int^t_{-\infty}
                            e^{-\lambda(t-s)}  \left\{
                             \left[
                             \varphi(s)-{ \tilde\varphi_{f} }(s)
                             \right] ds
                            + \frac{dY(s)} {2\left| \alpha \right|}
                                             \right \},
      \end{equation}
where the optimal Kalman gain ${ \tilde\Gamma}$ can be found numerically.
In order to solve this equation, we transform it into differential form as
      \begin{eqnarray}
       d{ \tilde\varphi_{f} }(t) &=& - \lambda { \tilde\varphi_{f} }(t) dt
                           + { \tilde\Gamma} \left\{ \left[
                             \varphi(t)-{ \tilde\varphi_{f} }(t)
                             \right] dt
                          + \frac{dY(t)} {2\left| \alpha \right|} \right \}
       \nonumber \\
                      &=& - (\lambda+{ \tilde\Gamma}) { \tilde\varphi_{f} }(t) dt
                           + { \tilde\Gamma} \left[
                             \varphi(t) dt
                          + \frac{dY(t)} {2\left| \alpha \right|}
                               \right].
      \end{eqnarray}
Then we transform it into integral form again,
      \begin{equation}
       { \tilde\varphi_{f} }(t) = { \tilde\Gamma} \int^t_{-\infty}
                          e^{-(\lambda+{ \tilde\Gamma})(t-s)}
                             \left[
                             \varphi(s) ds
                          + \frac{dY(s)} {2\left| \alpha \right|}
                                           \right ].
      \end{equation}
From this, we can see that the noise term $dY(t)$ is low-pass filtered with a cutoff (angular) frequency $(\lambda+{ \tilde\Gamma})$. Thus we could consider that the bandwidth of the estimator is $(\lambda+{ \tilde\Gamma})$.

We calculate the mean square error ${ \tilde\sigma_{f} }^2$ as,
      \begin{eqnarray}
        { \tilde\sigma_{f} }^2 &=&
                        \left<
                           \left[ \varphi(t)-{ \tilde\varphi_{f} }(t) \right]^2
                       \right>
                   \nonumber \\
               &=& \left< \left\{
                     { \tilde\Gamma} \int^t_{-\infty}
                          e^{-(\lambda+{ \tilde\Gamma})(t-s)}
                             \left[
                              \frac{\lambda+{ \tilde\Gamma}}{{ \tilde\Gamma}}\varphi(t)
                              -\varphi(s)
                             \right] ds
                   -{ \tilde\Gamma} \int^t_{-\infty}
                          e^{-(\lambda+{ \tilde\Gamma})(t-s)}
                          \frac{dY(s)} {2\left| \alpha \right|}
                  \right\}^2\right>
                     \nonumber \\
                &=&
                  { \tilde\Gamma}^2 \int^t_{-\infty} ds_1 \int^t_{-\infty} ds_2
                          e^{-(\lambda+{ \tilde\Gamma})(2t-s_1-s_2)}
                             \left<
                             \left[
                              \frac{\lambda+{ \tilde\Gamma}}{{ \tilde\Gamma}}\varphi(t)
                              -\varphi(s_1)
                             \right]
                             \left[
                              \frac{\lambda+{ \tilde\Gamma}}{{ \tilde\Gamma}}\varphi(t)
                              -\varphi(s_2)
                             \right]
                             \right>
                    \nonumber \\
                   & & +\frac{{\tilde\Gamma}^2}{4|\alpha|^2}
                       \int^t_{-\infty} \int^t_{-\infty}
                          e^{-(\lambda+{ \tilde\Gamma})(2t-s_1-s_2)}
                       \left \langle dY(s_1) dY(s_2)\right \rangle.
       \label{eq:sigmafFB}
      \end{eqnarray}
Here $\left<\varphi(s)\varphi(s')\right>$ is calculated from Eq.~\ref{eq:OUnoise} as,
      \begin{equation}
       \left<\varphi(s)\varphi(s')\right>=\frac{\kappa}{2\lambda}e^{-\lambda|s-s'|}.
       \label{eq:OUAutoCF}
      \end{equation}
Now ${ \tilde\sigma_{f} }^2$ can be given from Eqs. \ref{eq:dYAutoCF}, \ref{eq:ACFXpm}, \ref{eq:sigmafFB} and \ref{eq:OUAutoCF},
      \begin{eqnarray}
        { \tilde\sigma_{f} }^2 &=& \frac{\kappa}{2(\lambda+{ \tilde\Gamma})}
                   + \frac{1}{8|\alpha|^2} \frac{{ \tilde\Gamma}^2}{{ \tilde\Gamma}+\lambda}
             \nonumber \\
               & & \times \left\{
                          1+{ \tilde\sigma_{f} }^2(R_{+}-1)
                            \frac{h_{+}}{h_{+}+1}
                           + (1-{ \tilde\sigma_{f} }^2) (R_{-}-1)
                            \frac{h_{-}}{h_{-}+1}
                          \right\},
       \label{eq:sigmafFB2} \\
        h_{\pm}&=& \frac{ (1\mp x) \Delta \Omega_{0}} {\lambda+{ \tilde\Gamma}}.
      \end{eqnarray}
The effect of finite-bandwidth squeezing is characterized by $h_{\pm}$, which is the ratio of the squeezing (anti-squeezing) bandwidth $(1\mp x) \Delta \Omega_{0}$ to the estimator bandwidth $(\lambda+{ \tilde\Gamma} )$.
In the limit $\Delta\Omega_{0} \rightarrow \infty$ we find from Eq.~\ref{eq:sigmafFB2} that ${ \tilde\sigma_{f}^2}$
obtains its minimum value of $\sigma_f^2$ when $\tilde\Gamma = \Gamma$, as expected.

The optimal smoothed estimate with { finite-bandwidth }squeezing ${ \tilde\varphi_{s} }(t)$ is similarly
found to be
      \begin{equation}
        { \tilde\varphi_{s} }(t) = \left( 2\lambda+{ \tilde\Gamma} \right)
                        \int^\infty_{t} e^{-(\lambda+{ \tilde\Gamma})(s-t)}
                                               { \tilde\varphi_{f} }(s) ds,
      \end{equation}
 and the mean square error of the smoothed estimate is
      \begin{eqnarray}
        { \tilde\sigma_{s} }^2 &=& \frac{\kappa(2\lambda^2+{ \tilde\Gamma}^2+2\lambda { \tilde\Gamma})}
                             {4(\lambda+{ \tilde\Gamma})^3}
                      + \frac{{ \tilde\Gamma}^2(2\lambda+{ \tilde\Gamma})^2}
                             {16|\alpha|^2 (\lambda+{ \tilde\Gamma})^3}
         \nonumber \\
               & & \times \left\{
                          1+{ \tilde\sigma_{f} }^2(R_{+}-1)
                             \frac{h_{+}(h_{+}+2)}{(h_{+}+1)^2}
                           +(1-{ \tilde\sigma_{f} }^2) (R_{-}-1)
                             \frac{h_{-}(h_{-}+2)}{(h_{-}+1)^2}
                          \right\}.
       \label{eq:sigmasFB}
      \end{eqnarray}
Similarly to that for the filtered estimate, the effect of { finite-bandwidth }squeezing is characterized by $h_{\pm}$.
 Again, in the limit $\Delta\Omega_{0} \rightarrow \infty$ we find from Eq.~\ref{eq:sigmasFB} that ${ \tilde\sigma_{s}^2}$
obtains its minimum value of $\sigma_s^2$ when $\tilde\Gamma = \Gamma$.

\section*{S7 \ Effective squeezing bandwidth and photon flux}

In our experiment, we use narrow optical sidebands as our signal.
 The frequency range of squeezing which is exploited, in the vicinity of the sideband frequency, is thus much narrower
than the OPO's bandwidth.
In other words, most of the photon flux produced by the squeezing plays no part in the current experiment.
 Counting the entire photon flux of the beam would grossly
misrepresent the photon flux resource used by our phase tracking algorithm. To reasonably represent the
photons in the squeezing (in addition to those in the coherent field) we must consider the photons in a relatively
narrow band in the vicinity of the sideband frequency. That is, we must define the \textit{effective} squeezing (and
anti-squeezing) bandwidths.

Once again we model the spectrum of the squeezing and anti-squeezing by Eq.~\ref{eq:ROmega}, with
 $\Omega=0$ denoting the sideband frequency.
As before, $R_{\pm}$ denotes the squeezing (anti-squeezing) levels at the sideband frequency,
but now we replace $\Delta\Omega_0$ by an effective bandwidth $\Delta\Omega_{\rm eff}$ to model narrow-band squeezing.
The effective bandwidth is the narrowest bandwidth of squeezing that could have been used without substantially
affecting the experimental results obtained. A reasonable guess would be to set the effective bandwidth
somewhat wider than the estimator bandwidth.
This is because the squeezing within the estimator bandwidth can effectively enhance the estimate, whereas the squeezing outside of the estimator bandwidth is mostly wasted.
 Indeed, it turns out that defining $\Delta\Omega_{\rm  eff}= 2 (\lambda+\Gamma)$ is appropriate in the sense that ${ \tilde\sigma_{s} }^2$ is little different from $\sigma_s^2$ in the regime of our experiment.

From now on, we restrict to the experimental parameters as shown in Fig.~4 of the main text.
The squared coherent amplitude $|\alpha|^2$ ranges from $10^6$ to $10^7 s^{-1}$, and the other parameters are fixed as $\kappa= 1.9 \times10^4$ rad/s, $\lambda= 5.9\times10^4$ rad/s, $R_{-}=0.479$, and $R_{+}=3.09$ (see the main text).
The normalized pump amplitude $x$ is calculated as $x=0.33$ from Eq.~\ref{eq:x}.
 In Fig.~S7, we plot the mean square error of the phase estimate as a function of $|\alpha|^2$ (remember that $\Gamma$ is a function of $|\alpha|^2$ in Eq.~\ref{eq:KalmangainG}, and hence $\Delta\Omega_{\rm  eff}$ is too).
The green line indicates the mean square error $\sigma_s^2$ for broadband squeezing ($\Delta \Omega_{\rm  eff} \rightarrow \infty$).
The red line is the mean square error ${ \tilde\sigma_{s} }^2$ with $\Delta\Omega_{\rm  eff}= 2 (\lambda+\Gamma)$.
The difference between $\sigma_s^2$ and ${ \tilde\sigma_{s} }^2$ is less than 3\%, which is smaller than the experimental error bars which are $\sim 10\%$.

By setting $\Delta \Omega_{\rm  eff}=2(\lambda+\Gamma)$, we can now calculate photon flux of the squeezing
$\mathcal{N}^{\rm {} sq}_{\rm  eff}$ from Eq.~\ref{eq:Nsq}, and obtain the total effective photon flux $\mathcal{N}_{\rm  eff}= |\alpha|^2 + \mathcal{N}^{\rm {} sq}_{\rm  eff}$.
The contribution of $\mathcal{N}^{\rm {} sq}_{\rm eff}$ to the total photon flux is at most 7\% under our experimental conditions.
Figure S8 shows the mean square error versus the effective photon flux $\mathcal{N}_{\rm  eff}$ under the same conditions as Fig.~4 in the main text.
We also replot the experimental data from that figure.
To make an absolutely fair comparison, we correct the experimental mean square errors for broadband squeezing to the expected (i.e.\ slightly larger) values for the effective bandwidth.
The key point is that even taking into account the photon flux due to squeezing in the effective bandwidth around the sidebands, the accuracy of our phase estimates surpasses the ideal (lossless) coherent-state limit as a function of total photon flux over the whole range of the experiment.
\begin{figure}[h]
 \begin{center}
  \begin{tabular}{cc}
    \begin{minipage}{0.47\hsize}
      \begin{center}
       \includegraphics[width=75mm,clip]{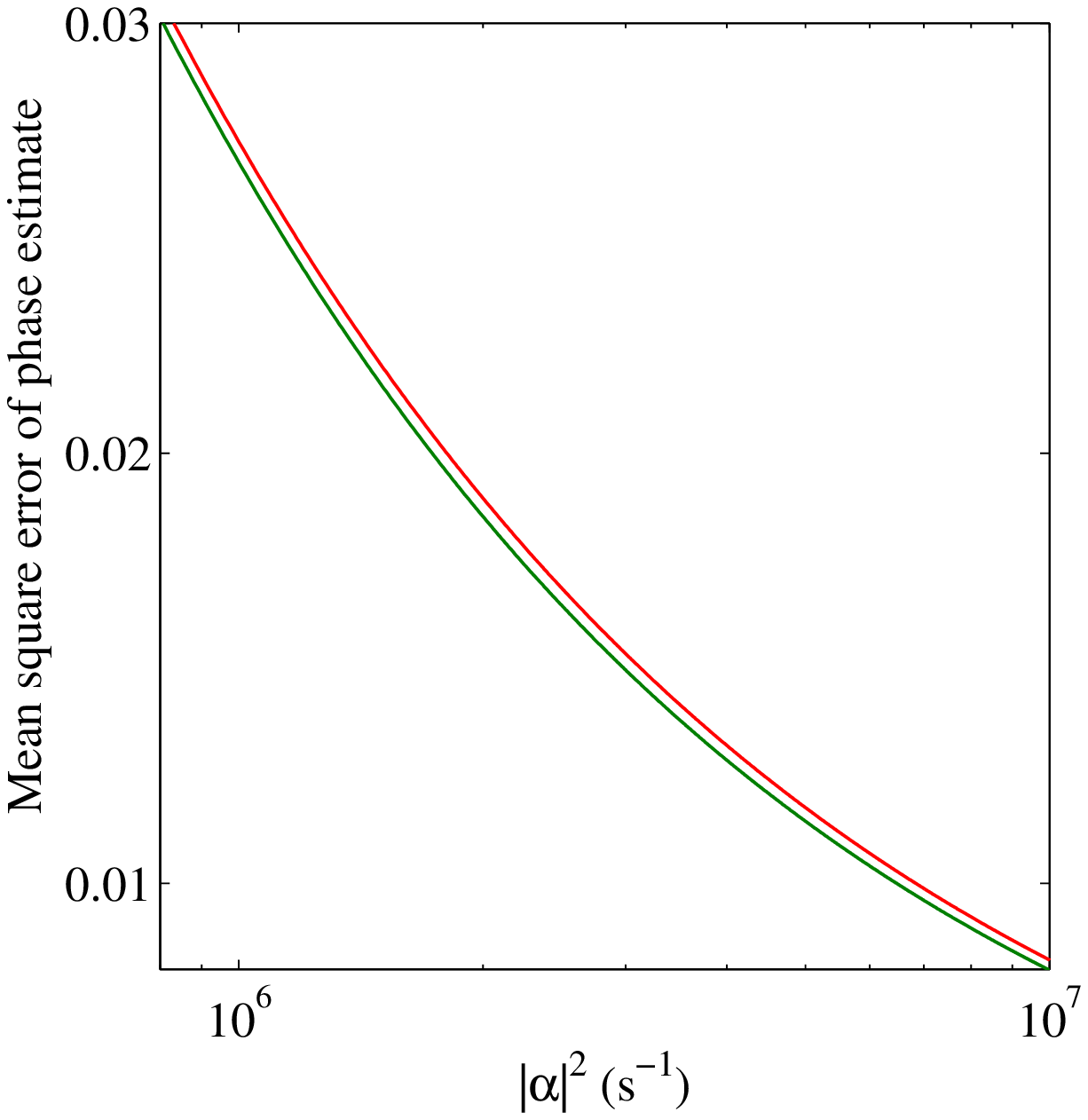}
       \caption{Theoretical curves for the mean square error in the smoothed
                estimate versus the photon flux $|\alpha|^2$ due to the
                coherent amplitude.
                Experimental conditions ($\kappa$, $\lambda$, and $R_{\pm}$)
                are same as Fig.~4 in the main text.
                The green line is $\sigma_{s}^2$ for broadband
                squeezing (as in the experiment). The red line is
                ${ \tilde\sigma_{s} }^2$
                for finite-bandwidth squeezing with
                $\Delta\Omega_{\rm  eff}= 2(\lambda+\Gamma)$, where $\Gamma$
                is the feedback gain.}
      \end{center}
    \end{minipage}
&
    \begin{minipage}{0.47\hsize}
      \begin{center}
      \includegraphics[width=75mm]{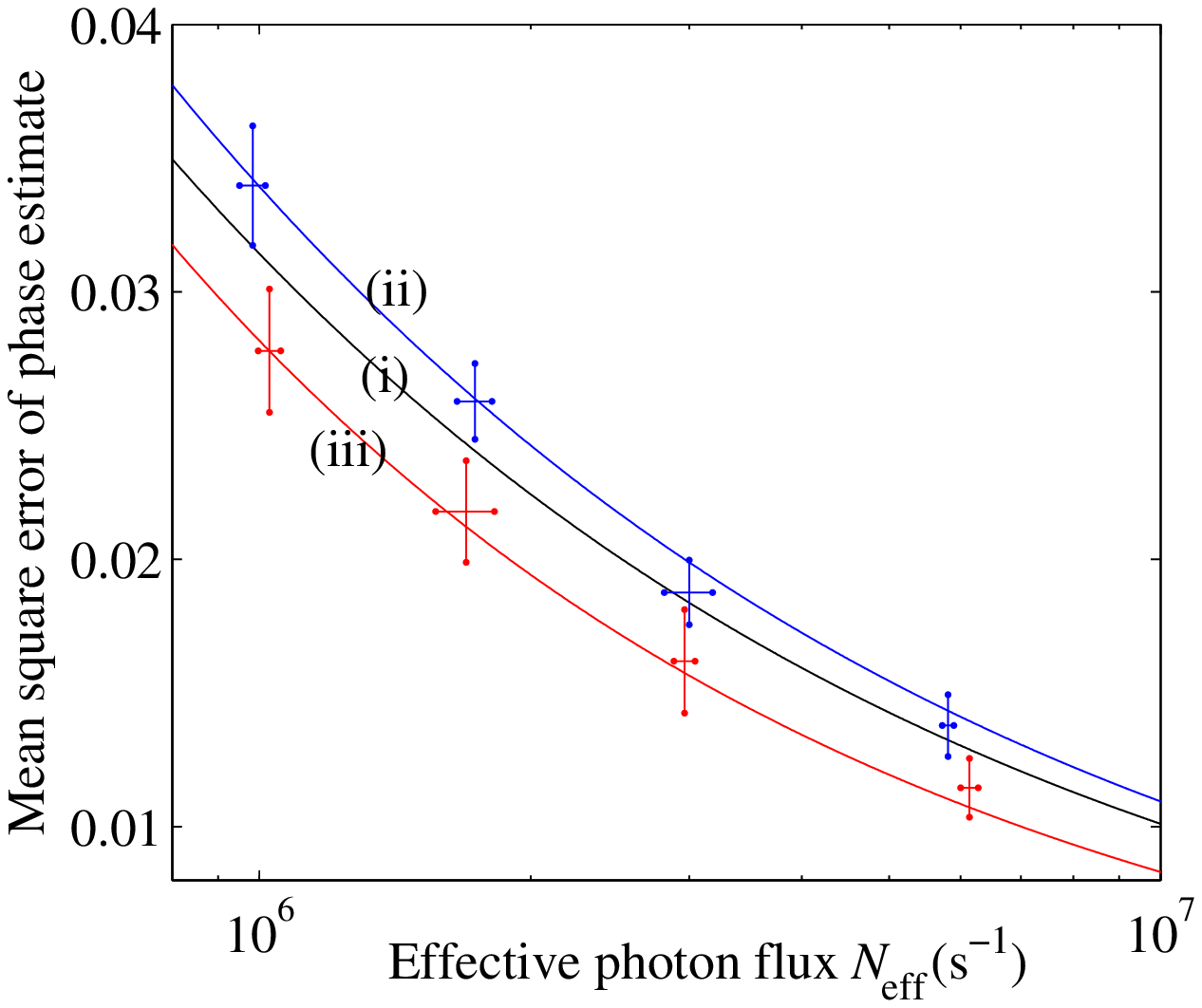}
        \caption{Mean square error of the smoothed estimate versus total
                 effective photon flux $\mathcal{N}_{\rm  eff}$.
                 Parameters are as in Fig.~S7.
                 Blue and red crosses are experimental data for
                 coherent and squeezed beams respectively. The latter are
                 corrected to the expected values for
                 squeezing with bandwidth $\Delta\Omega_{\rm  eff}$.
                 Trace (i) is the theoretical coherent-state limit. Trace (ii)
                 is the predicted (i.e. with experimental imperfections)
                 curve for coherent beams. Trace (iii) is the predicted curve
                 for { finite-bandwidth} squeezed beams including all
                 experimental imperfections.}
      \end{center}
    \end{minipage}
  \end{tabular}
 \end{center}
\end{figure}

\end{document}